# How do dataset characteristics affect the performance of propensity score methods and regression for controlling confounding in observational studies? A simulation study.

J. Wilkinson[*1], M. A. Mamas[2], E. Kontopantelis[3]

[1] Centre for Biostatistics, Manchester Academic Health Science Centre, Faculty of Biology, Medicine, and Health, University of Manchester, Rm 1.307 Jean McFarlane Building, University Place, Oxford Road, Manchester, M13 9PL, UK. jack.wilkinson@manchester.ac.uk

[2] Keele Cardiovascular Research Group, Centre for Prognosis Research, Keele University, Keele, UK.

[3] Division of Informatics, Imaging & Data Sciences, University of Manchester, Manchester, UK.

# Abstract

## Background

In observational studies, researchers must select a method to control for confounding. Options include propensity score methods and regression. It remains unclear how dataset characteristics (size, overlap in propensity scores, exposure prevalence) influence the relative performance of the methods, making it difficult to select the best method for a particular dataset. Given the increasing availability of large electronic health resources, performance for the analysis of big data is of particular interest.

## Methods

A simulation study to evaluate the role of dataset characteristics on the performance of several propensity score methods (followed by logistic regression), compared to multiple logistic regression, for estimating a marginal odds ratio in the presence of confounding. Outcomes were simulated from logistic and complementary log-log models, and size, overlap in propensity scores, and prevalence of the exposure were varied.

## Results

Multiple regression showed poor coverage for small sample sizes (100), but with large sample sizes (10,000 or above) it was more robust to imbalance in propensity scores and low exposure prevalence than were propensity score methods. Propensity score methods frequently displayed suboptimal coverage, particularly as overlap in propensity scores decreased. Problems associated with lack of overlap were exacerbated at larger sample sizes. Power of matching methods was particularly affected by lack of overlap, low prevalence of exposure, and small sample size. Performance of inverse probability of treatment weighting depended heavily on dataset characteristics, with poor coverage and bias with reduced overlap. The advantage of regression for large data size became less clear in the sensitivity analysis with a complementary log-log outcome generation mechanism and with unmeasured confounding, with superior bias and error but lower coverage than nearest neighbour and 1-to1 propensity score matching.

## Conclusions

Matching on the propensity score performed better in very small samples, but the performance of multiple regression was comparable in sample sizes of 1,000 and became increasingly superior as sample sizes increased. In contemporary large observation studies, of national registries or primary care electronic health records, multiple regression estimation is predominantly the optimal choice, both in terms of simplicity and performance.

# Introduction

Observational studies employing large, routinely collected datasets are now commonplace in the health sciences, exploiting new opportunities to study the effects of treatments or exposures in representative cohorts. A key concern in observational studies is how to address confounding, in order to permit the estimation of the effect of the exposure on an outcome. [1, 2] Researchers frequently use regression or propensity scores (PS) for this purpose, with the latter increasing in popularity in the past few decades [3].

The popularity of PS methods for observational health data can be attributed to several attractive features. For example, they offer intuitive checks for balance between groups which are not possible using regression methods [2, 4]. Additionally, they can be formulated without reference to the outcome. This might reduce bias arising from "p-hacking" (when analyses are selected on the basis of the results they produce) [5], because the impact on the estimated treatment effect is not known during development of the PS model [2]. Another advantage is the fact that regression methods implicitly but heavily rely on extrapolation when exposed and unexposed individuals have very different confounder distributions [6, 7]. This is more explicit for PS methods, because it manifests in the form of highly variable inverse probability weights or a lack of good matches. It is also appropriately reflected by reduced certainty in the estimated exposure effect [6, 7]. A final reason may lie in the fact that PS methods, particularly when PS are used for matching, are frequently described as "emulating" a randomised controlled trial. However, it might not be clear to those using this phrase that the success of this emulation depends on all confounders being included in the estimation of the PS.

The variety of available methods for handling confounding in observational studies creates a challenge for the applied health researcher, who must select the best analytic approach for the particular study at hand. For example, PS matching excludes some data from the analysis, and so its performance might depend on factors such as the study size and the proportion of individuals who are exposed. In addition, PS methods were developed in an era preceding the widespread availability of large health datasets, and evaluations of their performance have generally not considered large sample sizes. Evidence is therefore lacking on the relative performance of these methods for the analysis of big data. Previous studies have compared the results obtained by applying different methods [8], or have used simulation without investigating the impact of dataset characteristics on performance [9, 10]. We therefore conducted a comprehensive simulation study to evaluate commonly used approaches for confounding control, and to investigate the factors affecting their performance, with the aim of providing guidance for health researchers. We considered the roles of data size, imbalance in confounding variables, and the relative number of exposed to unexposed individuals.

# Methods

Methodological details are provided in the supplementary file.

## Propensity score

In a comparison of an exposed with a control group, PS can be estimated using multiple logistic regression where the binary outcome denotes membership of an exposed or a comparator group.

Covariates hypothesised to be associated with the outcome should be included in the multiple logistic regression model. The PS is the predicted probability $p$ of exposure. In practice, the regression coefficients must be estimated and hence there is uncertainty in $p$ that is not usually accounted for in the process (although it is possible to do so, e.g. [7, 11]). Following estimation, the next concern is to verify that these are balanced across the two groups [2]. Finally, PS are incorporated into the analysis using one of several approaches.

Stratification using the PS has been found to perform poorly, and so we decided not to consider it in our simulation study [12, 13]. Regressing the outcome on exposure and PS has been found to be the most commonly used approach in reviews of practice [14, 15]. Inverse probability of treatment weighting (IPTW) is an alternative, often used approach [16]. However, PS matching has gained a reputation as the best PS method for removing baseline imbalance and is widely used[17-19]. The most common version is one-to-one matching without replacement [2], where each "case" in group A is matched to one "control" in group B, if the difference in their PS is below a predefined arbitrary threshold or "caliper" [20]. This approach entails sample size reductions, which can become extreme if there is great imbalance in baseline characteristics (resulting in largely non-overlapping PS distributions in the study groups), or if there is a large imbalance in group sizes. An alternative is one-to-one nearest neighbour matching with replacement, which will generally result in fewer observations being discarded compared to matching on a caliper, at the expense of greater discrepancies in PS distributions between groups.

## Simulation study

We conducted a simulation study to evaluate PS methods and covariate adjustment for confounding control in observational studies, and the dataset characteristics affecting their performance.

### Data generating model

To investigate the influence of data size, we considered sample sizes of 100, 1000, 10,000, and 100,000, to capture scenarios in which large databases are available for analysis. We also investigated various scenarios for the distribution of the exposed and comparator groups: equal group sizes, imbalanced group sizes, and substantially imbalanced group sizes. A third varying parameter was baseline imbalance for the covariates, which took on five different patterns, ranging from well-overlapping propensity scores to almost completely non-overlapping propensity scores for the two comparison groups. Figure 1 shows the PS distributions when for equal exposed and comparator group sizes.

The simulation was implemented in Stata v15.1 [21]. We used the *drawnorm* command to draw observations from multivariate normal distributions, which were dichotomised for some variables. The generated variables included: binary exposure $E$, binary covariate $X_1$, and continuous covariates $X_2$, $X_3$, $X_4$ and $X_5$. Correlations were set to be low between all variables except for two of the covariates and the exposure. Following that, the outcome $Y$ was generated using a logit model. However, as a sensitivity analysis, we generated $Y$ using a complementary log-log model, to ensure that performance of PS methods and regression was evaluated under more neutral conditions. In additional sensitivity analyses, we included an unmeasured confounder in the outcome generation mechanism (a continuous covariate $X_6$ which did not feature in the analytical models), for either the logit or complementary log-log model.

## Analyses

A total of 5 analytical approaches were evaluated. First, PS were calculated using the PSMATCH2 module.[22] During this step we performed nearest neighbour one-to-one matching on the PS with replacement. Next, the PS were used in four logistic regression models: 1) exposure and the PS as independent variables (*PS covariate*); 2) exposure as the only independent variable, with the number of times each observation appeared in the aforementioned nearest-neighbour-matched dataset as a frequency weight (*nearest neighbour matching*); 3) exposure as the only independent variable, following one-to-one matching without replacement, when absolute difference on the PS was below $10^{-2}$ (*Caliper matching*); 4) exposure as the only independent variable and the PS used as an inverse probability treatment weight (*IPTW*). Note that in this study, we use standard logistic regression following matching, rather than a version intended for matched data; we return to this point in the discussion. We also performed logistic regression with the exposure and all five covariates included as independent variables (not using the PS), followed by regression standardisation, as a fifth approach [23]. Standardisation is necessary so that regression targets the same quantity as PS approaches (see *Target of inference*). We used the margins command with the post option following logistic regression to achieve this and used the delta method to compute confidence intervals on the log odds scale.

## Target of inference

We evaluated the methods against the marginal odds ratio, which is a measure of the exposure effect at the population level. We calculated the marginal odds ratio for each simulated dataset, using the method described by Austin [9]. When there is no heterogeneity in treatment effect, caliper and nearest neighbour matching, IPTW, and multiple logistic regression with standardisation all estimate the marginal odds ratio [9, 23]. PS covariate actually targets a different quantity; the odds ratio conditional on the PS [7, 24, 25]. We include it here due to its popularity, and to compare to other methods.

## Performance measures

1,000 datasets were simulated for each scenario. We considered four performance measures: mean error, bias, coverage and power. Mean error is the mean of the absolute difference between the estimate and the true parameter. Bias is the mean difference between the estimate and the true parameter. Coverage is the proportion of 95% confidence intervals for the estimate, based on a normal approximation, that contain the true parameter. Finally, we calculated power by the proportion of iterations where the null was rejected when it was actually false. Although power as a metric can be problematic in the presence of bias, it is essential for a complete comparison. However, in order to obtain a more meaningful metric, power-related statistical significance was calculated one-sided (i.e. statistically greater than zero, rather than statistically different). We also evaluated model convergence. The other metrics were only computed when convergence for a particular method in a simulation setting was 25% or above, otherwise they were set to missing.

# Results

Figures 2 to 7 show the results of the main simulation study. Supplementary tables S 1-3 give the numerical results, including standard errors for the performance metrics. Results for the sensitivity analyses (neutral comparisons, and introduction of an unmeasured confounder) are shown in the Supplementary File. We describe the results for the main analysis below, noting where sensitivity analyses resulted in departures from the main study results.

### Convergence

As expected, convergence of all methods was adversely affected by reduced exposure prevalence, decreasing overlap in PS, and reduced sample size (Figure 2). All approaches converged infrequently at smaller sample sizes when there was little overlap in PS; IPTW and multiple logistic regression were most robust. With n=100,000, these two methods generally converged even when the PS distributions were not overlapping (scenario 5) and exposure prevalence was very low (5%), although use of a complementary log-log link for outcome generation adversely affected this behaviour (Supplementary File). Convergence for PS covariate was particularly affected by confounding (lack of PS overlap); convergence was actually reduced when there was little overlap for larger (n = 100,000) compared to smaller (n = 1,000, 10,000) sample sizes. This was also observed when comparing datasets of n=100,000 to n= 10,000 for caliper matching, and nearest neighbour matching when exposure prevalence was low (10%) or very low (5%).

### Bias and absolute error

Bias and absolute error were consistently low for multiple logistic regression compared to other methods (Figure 3, 4), although IPTW was less biased for n = 100, when exposure prevalence was very low (5%) and there was overlap in PS distributions. Both measures were affected by sample size, with bias and/or error in the presence of non-overlapping PS distributions actually becoming more pronounced with increasing data size for some methods. IPTW in particular had high bias and error when overlap was low and sample sizes were large. Caliper matching was consistently better than nearest neighbour matching. Despite targeting a different estimand, PS covariate fared relatively well, although performance broke down under challenging circumstances (combinations of low exposure prevalence, small sample size, and little overlap in PS).

### Power and coverage

In the main scenario, power was generally as high or higher than other methods for multiple logistic regression, although IPTW had higher power in several scenarios where PS distributions were non-overlapping (Figure 5). Coverage of IPTW was generally poor in these scenarios however (Figure 6), and performance was consistently inferior when both power and coverage were considered (Figure 7). Power of matching methods was greatly affected by lack of overlap in PS distributions. For caliper matching without replacement, when there is substantial imbalance there will tend to be few matches. Consequently, power when there was reduced overlap was sometimes superior for nearest neighbour matching. Additionally, sample size following matching is smaller when exposure prevalence is low, and this also affected power for matching compared to other methods, particularly when sample sizes were small to start off with. Coverage was decreased at n = 100,000 for the matching methods compared to sample sizes of 1,000 and 10,000 when there was imbalance in PS. Power of PS covariate compared to other PS methods was only favourable when there was considerable overlap in PS distributions or 50% exposure prevalence; coverage was frequently but not consistently superior. However, when a complementary log-log link was used, coverage was sometimes inferior for logistic regression compared to 1-to-1 and nearest neighbour propensity score matching, specifically with larger data size and modest or high imbalance in PS. This was exacerbated when an unmeasured confounder was added to the complementary log-log link scenario. Power tended to be poor for matching methods in these scenarios (both with and without unmeasured confounding). Overall, when considering power and coverage as a composite, 1-to-1 and nearest neighbour matching appeared superior to regression for large data size with a complementary log-log outcome in the presence of unmeasured confounding. However, logistic regression remained superior in both absolute error and mean bias across all these scenarios.

## Discussion

We conducted a simulation study to compare several approaches to estimate a marginal odds ratio in the presence of confounding and to investigate how dataset characteristics influenced performance. In the main study, multiple logistic regression followed by standardisation was consistently superior to PS approaches, although coverage still fell short of the advertised level for very small sample sizes (n=100) or for sample sizes of 1,000 when there was limited overlap in PS distributions. It was observed to be quite robust to imbalance in PS for large sample sizes, even when exposure prevalence was very low.

We explored whether simulating from a logistic model conferred an advantage to multiple logistic regression, using an alternative outcome generation approach (complementary log-log link). We also introduced an unmeasured confounder in additional sensitivity analyses. The combination of a complementary log-log outcome generating model and introduction of an unmeasured confounder resulted in somewhat different results for large data sizes. While bias and absolute error remained lower for logistic regression, coverage became very poor at times, even when imbalance in PS was not severe. While matching methods had poor power for the combination of large data size, high imbalance in PS, and presence of unmeasured confounding, 1-to-1 and nearest neighbour matching appeared superior to regression when considering a composite of coverage and power. This was due to regression having a much smaller model standard error compared to matching in the large data scenario, since these two matching approaches discard data. This resulted in the regression confidence interval frequently failing to cover the true value, while the much wider matching confidence intervals would frequently span both the true value (hence, higher, but still suboptimal coverage) and the null (hence, lower power).

However, logistic regression remained the best performer in terms of absolute error and mean bias, across all these scenarios. In addition, the size of the true effect is relevant here, and the effect we modelled was relatively large for samples of 10,000 or 100,000, thus not allowing for discrimination in power in these scenarios (power was 100%). We a-priori decided on the modelled effect to allow for at least some discrimination in the results of the smaller samples, and we did not vary it across scenarios to allow for meaningful comparisons. However, a smaller modelled effect would greatly reduce power for the matching approaches (due to standard errors two to three times those of logistic regression), leading to logistic regression being the best performer in the cumulative of power and coverage, in line with what we observed for samples of 1,000.

Coverage of PS methods was frequently suboptimal, as has been previously observed for a non-null marginal odds ratio [9]. As anticipated, relative performance of PS methods depended on dataset characteristics. Ahead of the study, it was anticipated that matching using PS might be particularly affected by small sample size, imbalance in the PS distributions, and low exposure prevalence. While power was affected by these factors, overall performance of matching methods withstood these challenges better than IPTW. Matching methods also converged more frequently than did adjusting for the PS as a covariate in the presence of imbalance in PS distributions and large data sizes. Caliper matching was slightly preferred to nearest neighbour matching overall, although nearest neighbour did achieve superior power and coverage in some scenarios.

IPTW displayed poor performance as overlap in PS decreased. This suggests that IPTW is only suitable when there is substantial overlap in PS, highlighting the importance of examining distributions of the estimated PS [2, 26]. Vandersteedt and Daniels similarly found poor performance of IPTW when there was limited overlap in PS, up to a sample size of 1000; our results show that sample sizes considerably larger than this do not alleviate the problems [7]. However, we did not consider the use of stabilised or truncated weights [27], and it is unclear whether these would have led to improved performance. PS covariate does not estimate the marginal odds ratio but instead targets a different, and arguably unusual quantity – the odds ratio conditional on the PS. Performance against the marginal odds ratio was generally unacceptable, frequently falling short of the advertised coverage level when there was not high overlap in PS distributions and when there were small numbers of exposed participants, and for large data sizes balance did not guarantee appropriate coverage. PS covariate did not usually converge for larger data sizes when there was less overlap in PS distribution.

While we have considered the role of dataset characteristics in selecting a method for controlling confounding, there are several outstanding questions. One question is whether a paired or unpaired version of regression should be used after PS matching. There is uncertainty around this point, because people matched on their PS may nonetheless differ in terms of their covariate values. This question has been addressed in relation to continuous outcomes [28], but it remains to settle the issue in relation to binary outcomes. Suboptimal coverage against a non-null marginal odds ratio has been previously observed for several methods for analysing paired data following PS matching [9], and in the present study we also found this to hold when using an unpaired regression method. A direct comparison would be useful for future work.

By design, we did not include covariates in the regression models incorporating the PS (as a covariate, with IPTW, or following matching). Including covariates in the PS covariate approach results in a doubly-robust estimator, which offers valid inference in relation to some summary measures of the exposure effect (other than odds ratios) provided that one of the PS model and outcome model are correctly specified [29, 30]. Aside from this protection against misspecification, Vansteelandt and Davies found some power advantages when additionally adjusting for covariates in the outcome regression model compared to adjusting for the PS alone [7]. Consistent benefits of adjusting for covariates when using IPTW were not observed.

We have not considered the case where there is heterogeneity in the exposure effect across strata defined by the PS. When there is heterogeneity, different PS methods target different quantities, and might produce substantially different answers as a result [31], [32]. For example, matching estimates the exposure effect in the population corresponding to those who were exposed in the study, because the matching process produces a sample with similar PS distributions to the exposed group [32] (or rather, for caliper matching, similar to the exposed participants *for whom unexposed matches could be found* [31]). These considerations motivate the suggestion that the possibility of an interaction between the PS and exposure should be routinely examined [33].

## Conclusion

Researchers analysing observational data often face difficult analytical choices, while propensity score approaches are not easy to implement in large databases of electronic health records. Our results show how key features of a dataset (size, exposure prevalence, imbalance in propensity scores) affect the performance of several approaches aiming to address confounding. This study suggests that multiple logistic regression is relatively robust to low exposure prevalence and imbalance in PS, outside of very small sample sizes. For large sample sizes, multiple logistic regression was clearly the preferred method, especially in the main scenario, while PS methods performed poorly as imbalance in PS distributions increased, and this was not mitigated by large sample size or balanced group sizes. This highlights the importance of examining overlap in PS if these methods are to be used, but also suggests that their performance is worst when the problem they are intended to solve is most severe. Coverage of logistic regression was inferior to 1-to-1 and nearest neighbour propensity score matching methods in some large-data scenarios, however, particularly when a complementary log-log outcome generating model was used and either a) imbalance in PS was moderate or high, or b) an unmeasured confounder was introduced. However, this was driven by much larger standard errors in these two matching approaches, while logistic regression remained the best performer in absolute error and mean bias. In contemporary large observation studies, of national registries or primary care electronic health records, multiple regression estimation appears to be the optimal choice, both in terms of simplicity and performance.

**Figures**
*Figure 1: Simulated propensity score scenarios, when Pr(E=1)=0.5*
*Figure 2: Convergence (%)*
*Figure 3: Bias*
*Figure 4: Absolute error*
*Figure 5: Power (%)*
*Figure 6: Coverage (%)*
*Figure 7: Mean of coverage and power (%)*

*Figure 1: Simulated propensity score scenarios, when Pr(E=1)=0.5*

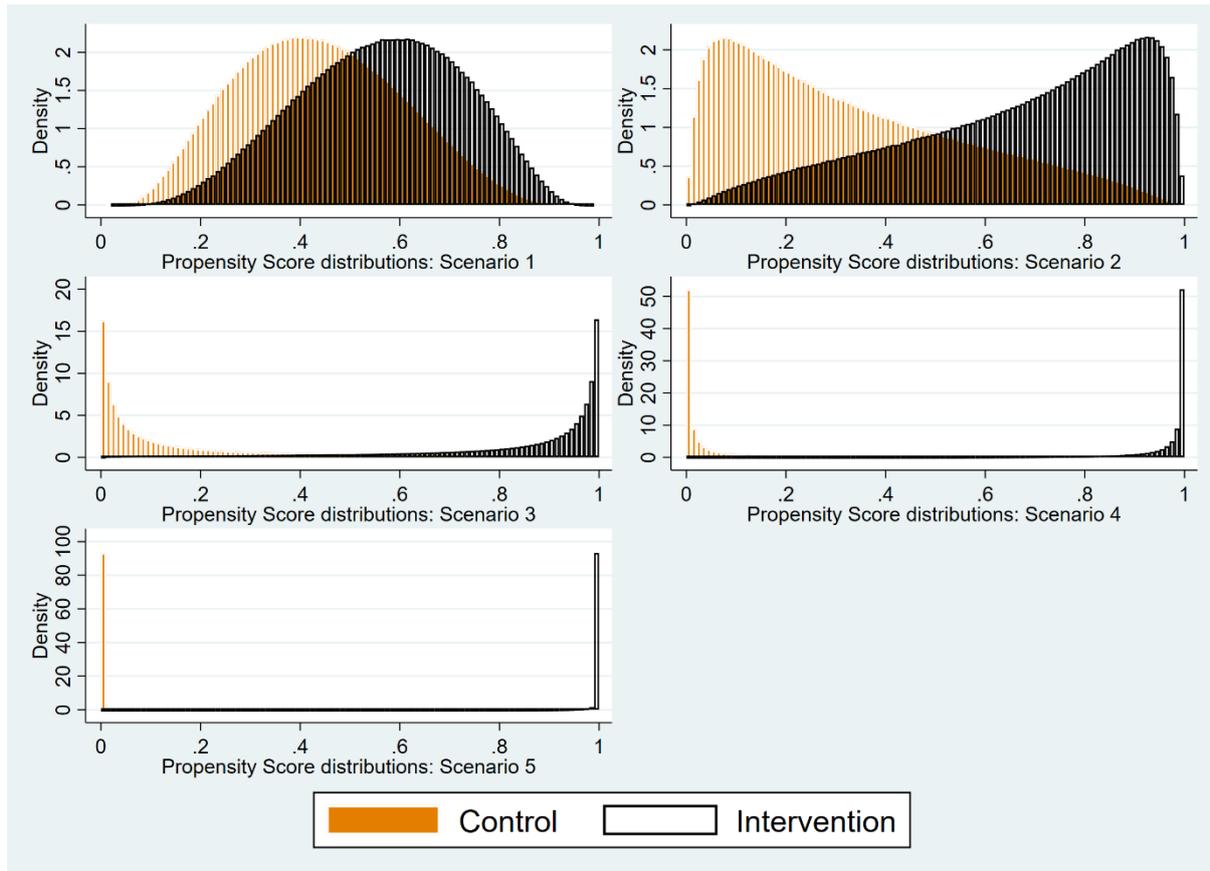

*Figure 2: Convergence (%)*

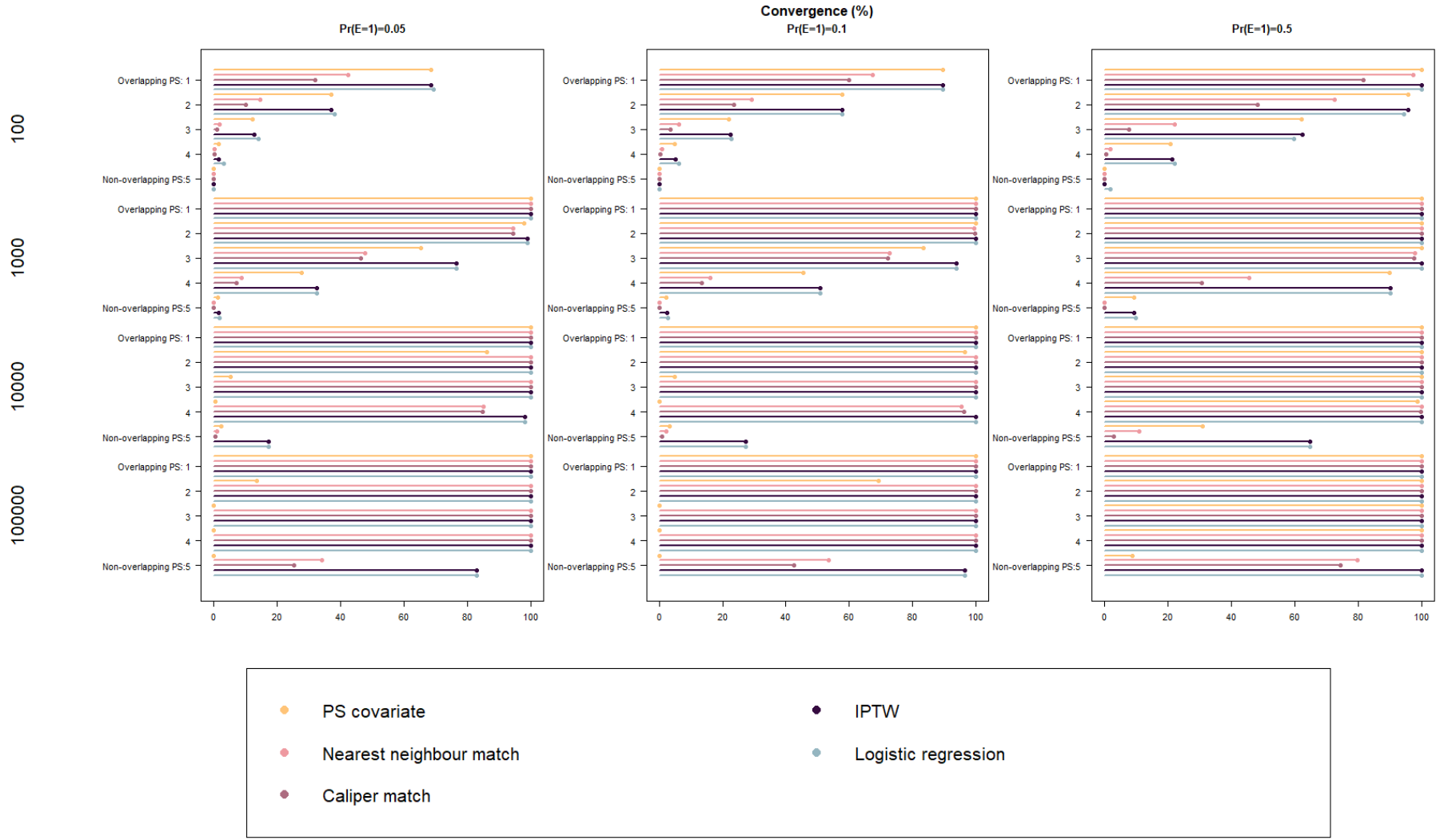

*Figure 3: Bias*

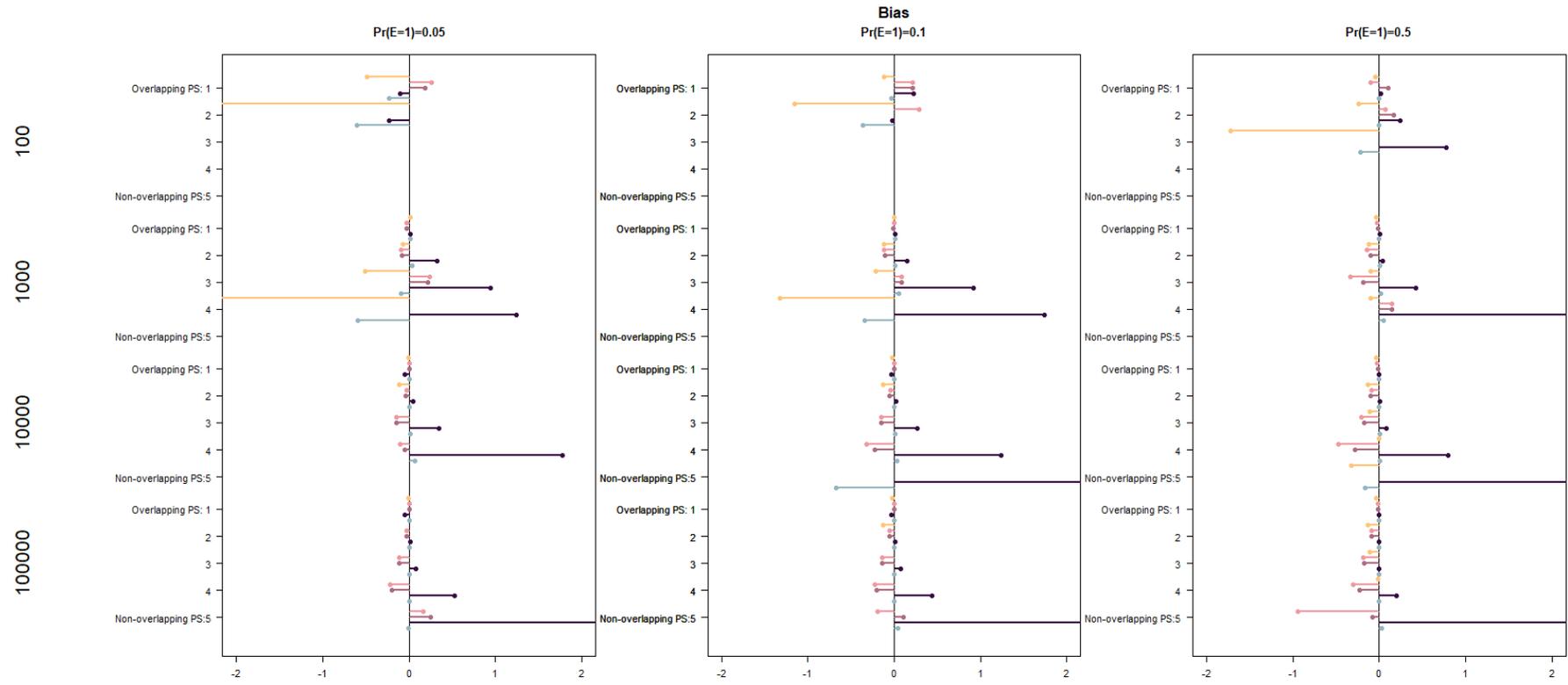

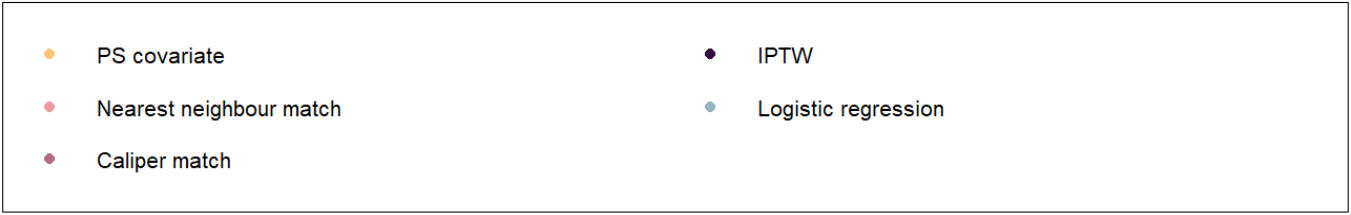

*Figure 4: Absolute error*

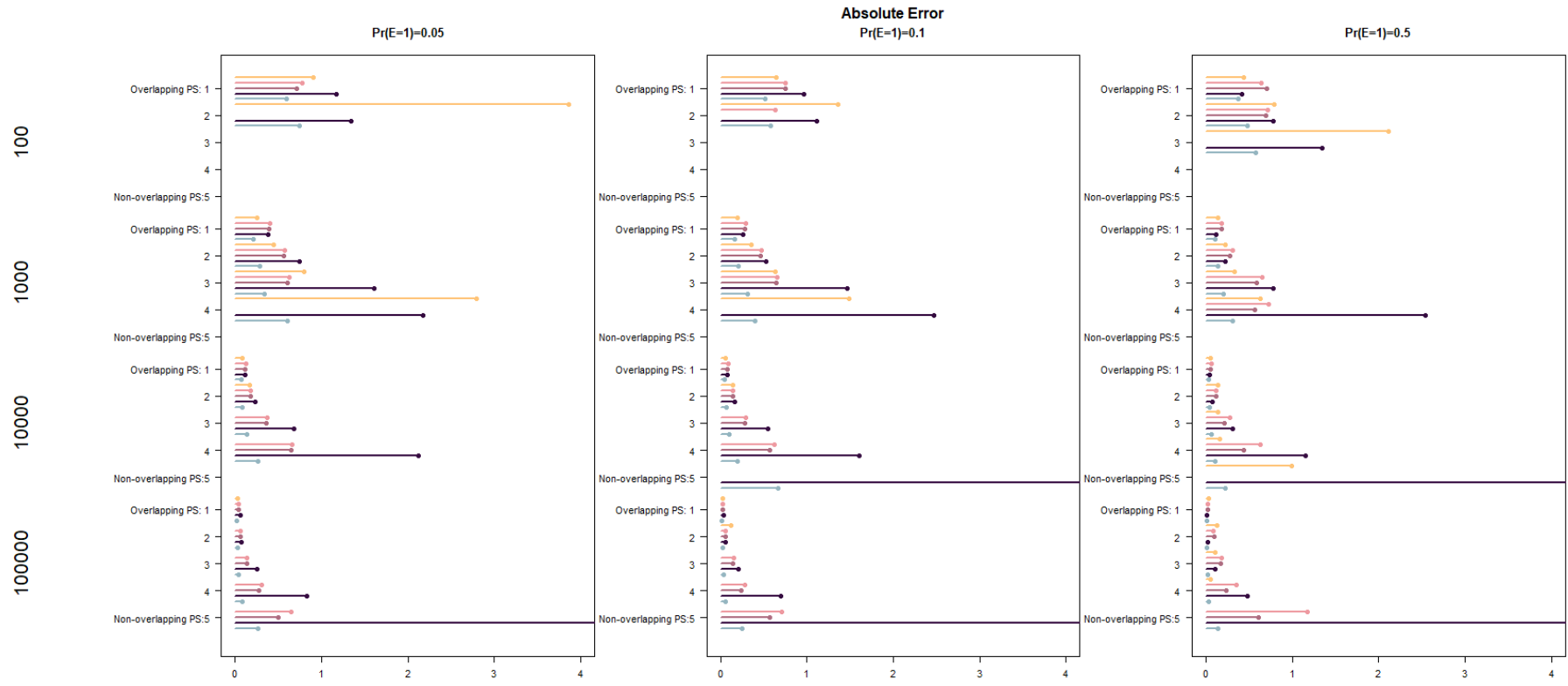
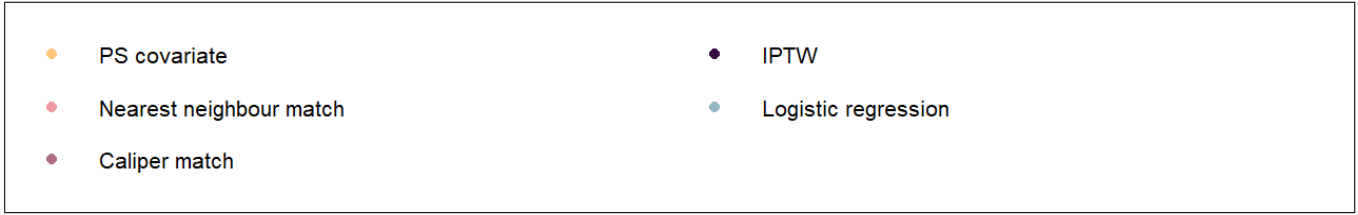

*Figure 5: Power (%)*

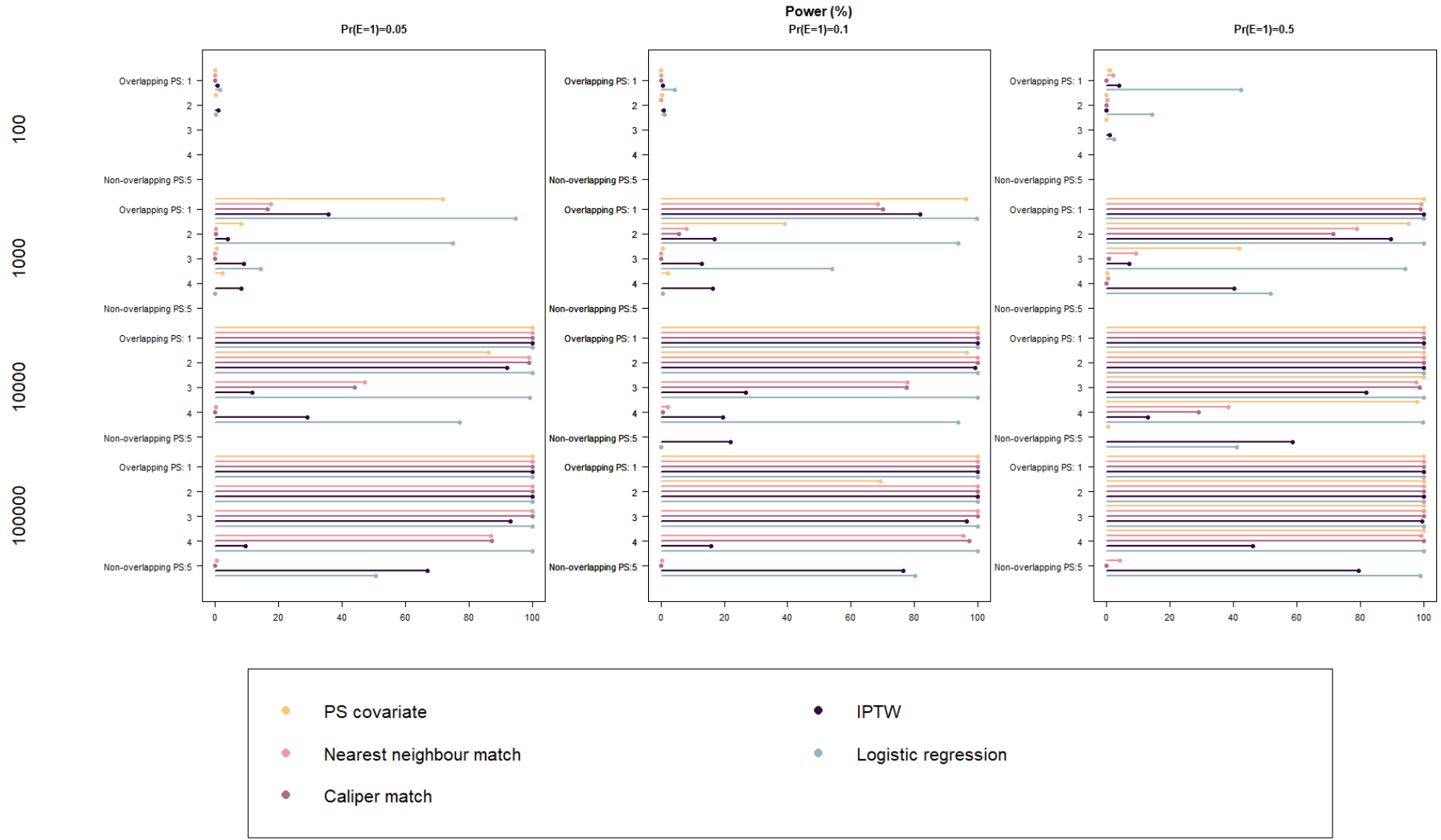

*Figure 6: Coverage (%)*

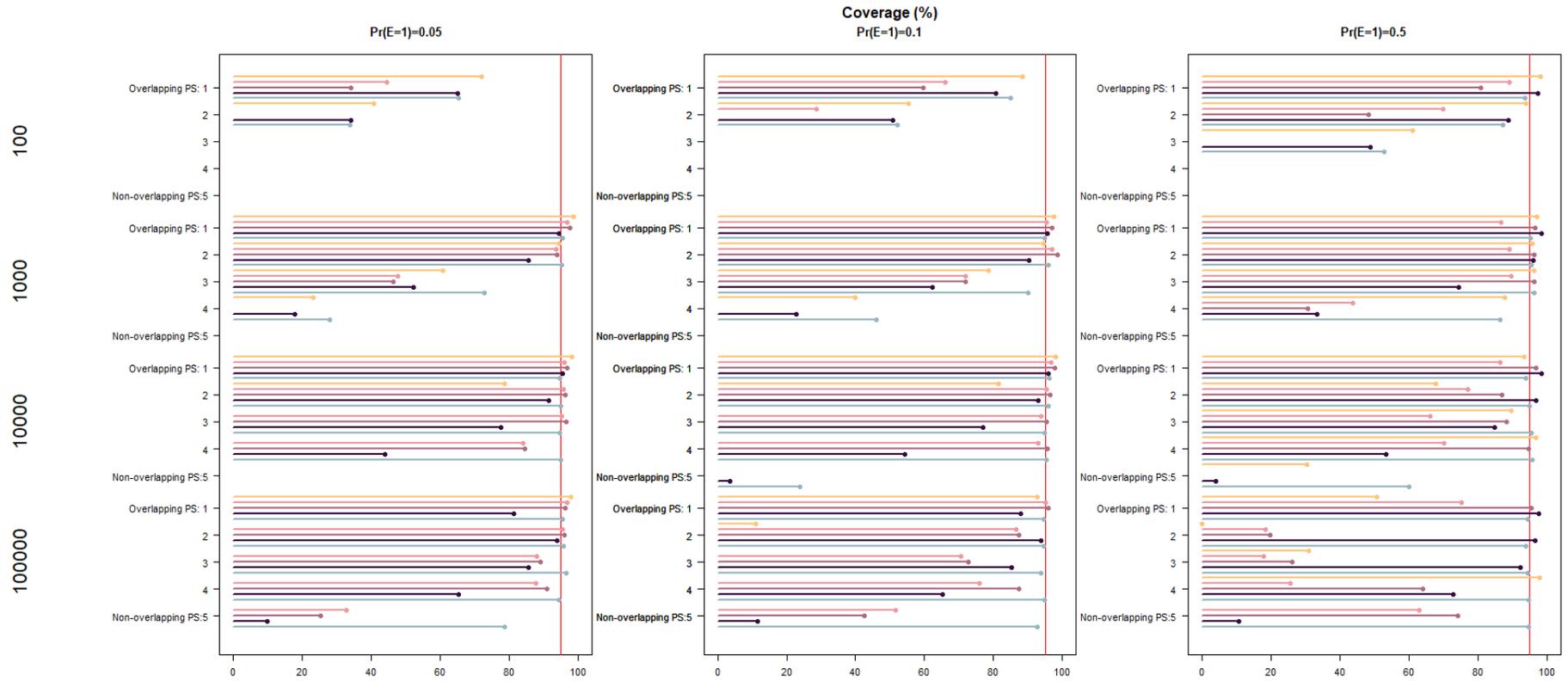

*Figure 7: Mean of coverage and power (%)*

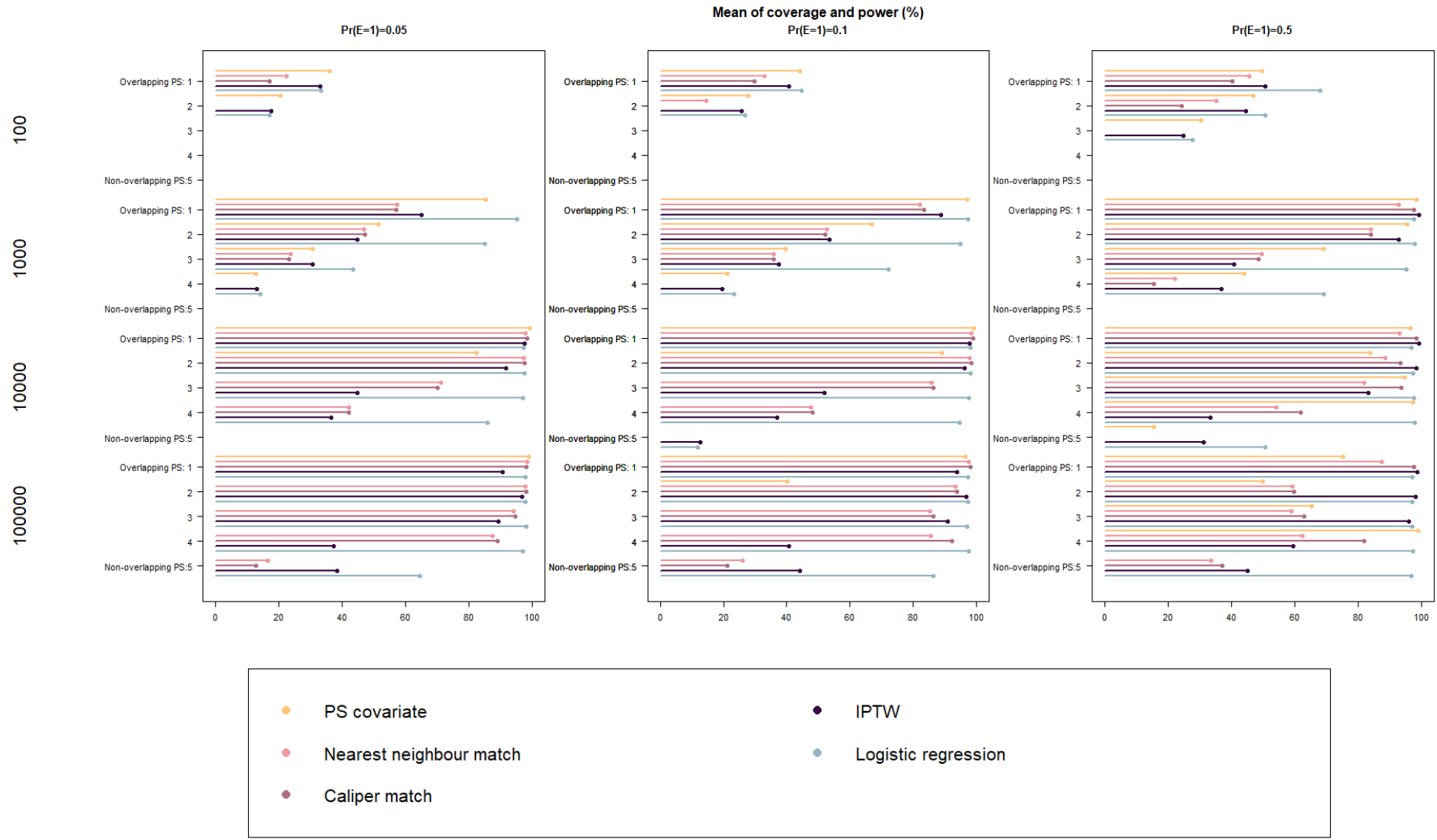

# Supplementary file for "How do dataset characteristics affect the performance of propensity score methods and regression for controlling confounding in observational studies? A simulation study."

## Methods

### Propensity score

In a comparison of an exposed with a control group, PS can be estimated using multiple logistic regression where the binary response variable *E* denotes membership of an exposed (*E* = 1) or a comparator (*E* = 0) group. We suppress a subscript *j* corresponding to the *j* = 1, …, n participants included in the study. Covariates $x_1, x_2, …, x_k$ hypothesised to be associated with the outcome should be included in the multiple logistic regression model. The PS is the predicted probability $p$ of exposure (*E*= 1):

$$Log\left(\frac{p}{1-p}\right) = \beta_0 + \beta_1 x_1 + \cdots + \beta_k x_k \quad (1)$$

In practice, coefficients $\beta_i$, *i =1,…,k* must be estimated and hence there is uncertainty in $p$ that is not usually accounted for in the process (although it is possible to do so, e.g. [1, 2]). Following estimation, the next concern is to verify that these are balanced across the two groups [3]. Finally, PS are incorporated into the analysis using one of several approaches.

Stratification using the PS has been found to perform poorly, and so we decided not to consider it in our simulation study [4, 5]. Regressing the outcome on exposure and PS has been found to be the most commonly used approach in reviews of practice [6, 7].

Inverse probability of treatment weighting (IPTW) is an alternative approach [8]. Weights corresponding to the inverse of the probability that the participant is included in their group are defined:

$$w = \begin{cases} 1/PS & if\ E = 1 \\ 1/(1-PS) & if\ E = 0 \end{cases} \quad (2)$$

Observations are then weighted accordingly in the analysis. PS matching has gained a reputation as the best PS method for removing baseline imbalance and is widely used[9-11]. The most common version is one-to-one matching without replacement [3], where each "case" in group A is matched to one "control" in group B, if the difference in their PS is below a predefined arbitrary threshold or "caliper" $\delta$, i.e. $|PS_A - PS_B| \leq \delta$ [12]. This approach entails sample size reductions, which can become extreme if there is great imbalance in baseline characteristics (resulting in largely non-overlapping PS distributions in the study groups), or if there is a large imbalance in group sizes. An alternative is one-to-one nearest neighbour matching with replacement, which will generally result in fewer observations being discarded compared to matching on a caliper, at the expense of greater discrepancies in PS distributions between groups.

# Simulation study

We conducted a simulation study to evaluate PS methods and covariate adjustment for confounding control in observational studies, and the dataset characteristics affecting their performance.

## Data generating model

To investigate the influence of data size, we considered sample sizes of 100, 1000, 10000, and 100000, to capture scenarios in which large databases are available for analysis. We also investigated the impact of the distribution of $E$: we considered equal group sizes $Pr(E = 1) = 0.5$, imbalanced group sizes $Pr(E = 1) = 0.1$, and substantially imbalanced group sizes $Pr(E = 1) = 0.05$. A third varying parameter was baseline imbalance for the covariates, which took on five different patterns, ranging from well-overlapping propensity scores to almost completely non-overlapping propensity scores for the two comparison groups. S Figure 1 shows the PS distributions when Pr(E=1)=0.5.

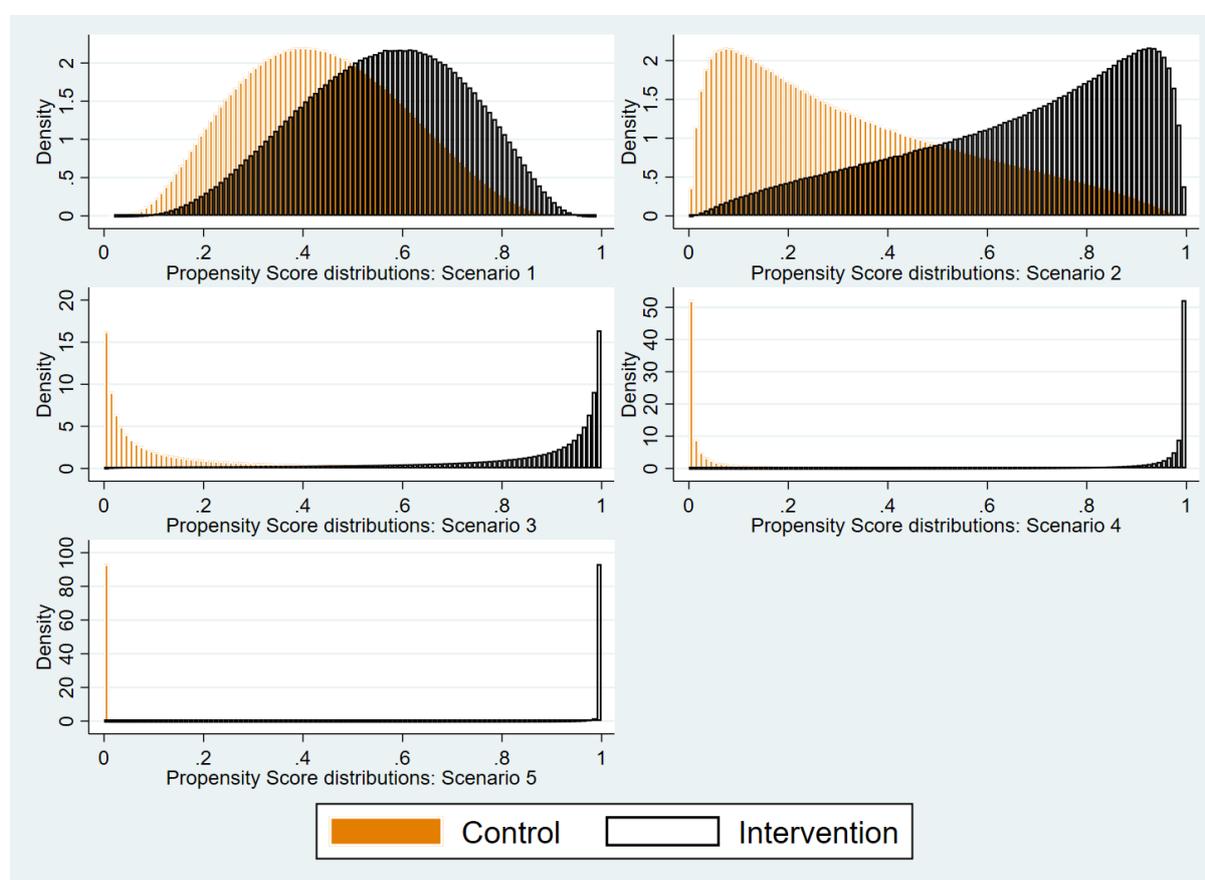

*S Figure 1: Simulated propensity score scenarios, when Pr(E=1)=0.5*

### Generation of covariates

The simulation was implemented in Stata v15.1 [13]. We used the *drawnorm* command to draw observations from multivariate normal distributions, which were dichotomised for some variables. The generated variables included: binary exposure $E$, binary covariate $X_1$ (with $Pr(X_1 = 1) = 0.5$), and continuous covariates $X_2, X_3, X_4$ and $X_5$ (with $X_i \sim N(0,1)$ for $i = 2,3,4,5$). Correlations were set to be low between all variables (≈0.1 Pearson's correlation) except for two of the covariates and the exposure ($corr_{Tetrachoric}(X_1, E) \approx 0.3$ and $corr_{Pearson}(X_2, E) \approx 0.5$). These associations were necessarily affected to a small extent when baseline differences were incorporated by shifting the distributions of the covariates for the $E = 1$ group. When PS distributions were simulated to be overlapping (S Figure 1, Scenario 1), all continuous covariates were edited for $E = 1$ so that the mean difference $E(X_i|E = 1) - E(X_i|E = 0) = 0.5$ for $i = 2,3,4,5$, while $Pr(X_1 = 1|E = 1)$ was set to 0.45. For scenarios 2 to 5 (S Figure 1), the mean difference was 1.0, 1.5, 2.0 and 3.0, respectively; and $Pr(X_1 = 1|E = 1)$ was 0.40, 0.35, 0.30 and 0.20, respectively. Note that providing exact distributions for continuous covariates when $E = 1$ is not straightforward due to the modelled correlations (e.g. the distribution for $X_2$ differs across $E$ regardless of the other simulation parameters).

*Outcome generation*

In the last step, the outcome $Y$ was generated:

$$logit(\pi) = -2.3 + ln(2)\, E + ln(1.3)\, X_1 + ln(1.5)\, X_2 + ln(6)\, X_3 + ln(3)\, X_4 + ln(1)\, X_5, \quad (3)$$
$$Y \sim \text{Bernoulli}(\pi).$$

$X_5$ is an instrumental variable; it is associated with exposure, but not with outcome.

*Alternative outcome generation approaches*

To examine the robustness of results to varying the outcome generation mechanism, we also used a complementary log-log link to generate the outcome $Y$:

$$log\{-log(1-\pi)\} = -2.3 + ln(2)\, E + ln(1.3)\, X_1 + ln(1.5)\, X_2 + ln(6)\, X_3 + ln(3)\, X_4 + ln(1)\, X_5, \quad (4)$$
$$Y \sim \text{Bernoulli}(\pi).$$

In additional sensitivity analyses we introduced an unmeasured confounder $X_6$, in each of the two link approaches, which was not included in any of the analytical models:

$$logit(\pi) = -2.3 + ln(2)\, E + ln(1.3)\, X_1 + ln(1.5)\, X_2 + ln(6)\, X_3 + ln(3)\, X_4 + ln(1)\, X_5 + ln(2)\, X_6, \quad (5)$$
$$Y \sim \text{Bernoulli}(\pi).$$

And

$$log\{-log(1-\pi)\} = -2.3 + ln(2)\, E + ln(1.3)\, X_1 + ln(1.5)\, X_2 + ln(6)\, X_3 + ln(3)\, X_4 + ln(1)\, X_5 + ln(2)\, X_6, \quad (6)$$
$$Y \sim \text{Bernoulli}(\pi).$$

Analyses

A total of 5 analytical approaches were evaluated. First, PS were calculated using the PSMATCH2 module.[14] During this step we performed nearest neighbour one-to-one matching on the PS with replacement. Next, the PS were used in four logistic regression models: 1) exposure and the PS as independent variables (*PS covariate*); 2) exposure as the only independent variable, with the number of times each observation appeared in the aforementioned nearest-neighbour-matched dataset as a frequency weight (*nearest neighbour matching*); 3) exposure as the only independent variable, following one-to-one matching without replacement, when absolute difference on the PS was below $10^{-2}$ (*Caliper matching*); 4) exposure as the only independent variable and the PS used as an inverse probability treatment weight (*IPTW*). Note that in this study, we use standard logistic regression following matching, rather than a version intended for matched data; we return to this point in the discussion. We also performed logistic regression with the exposure and all five covariates included as independent variables (not using the PS), followed by regression standardisation, as a fifth approach [15]. Standardisation is necessary so that regression targets the same quantity as PS approaches (see *Target of inference*). We used the margins command with the post option following logistic regression to achieve this, and used the delta method to compute confidence intervals on the log odds scale.

### Target of inference

We evaluated the methods against the marginal odds ratio, which is a measure of the exposure effect at the population level. We calculated the marginal odds ratio for each simulated dataset, using the method described by Austin [16]. When there is no heterogeneity in treatment effect, caliper and nearest neighbour matching, IPTW, and logistic regression with standardisation all estimate the marginal odds ratio [15, 16]. PS covariate actually targets a different quantity; the odds ratio conditional on the PS [1, 17, 18]. We include it here due to its popularity, and to compare to other methods.

### Performance measures

1000 datasets were simulated for each scenario. We considered four performance measures: mean error, bias, coverage and power. Mean error is the mean of the absolute difference between the estimate and the true parameter: $\frac{1}{1000}\sum_{i=1}^{1000}|z - \hat{z}_i|$ where $z$ is the true association and $\hat{z}_i$ is the estimate of the association. Bias is the mean difference between the estimate and the true parameter, or $\frac{1}{1000}\sum_{i=1}^{1000}(z - \hat{z}_i)$. Coverage is the proportion of 95% confidence intervals for the estimate, based on a Normal approximation, that contain the true parameter. Finally, we calculated power by the proportion of iterations where the null was rejected when it was actually false. Although power as a metric can be problematic in the presence of bias, it is essential for a complete comparison. However, in order to obtain a more meaningful metric, power-related statistical significance was calculated one-sided (i.e. statistically greater than zero, rather than statistically different). We also evaluated model convergence. We used default settings for convergence evaluation in Stata 15.1 MP [13]. The other metrics were only computed when convergence for a particular method in a simulation setting was 25% or above, otherwise they were set to missing.

Sensitivity 1: complementary log-log link outcome generation

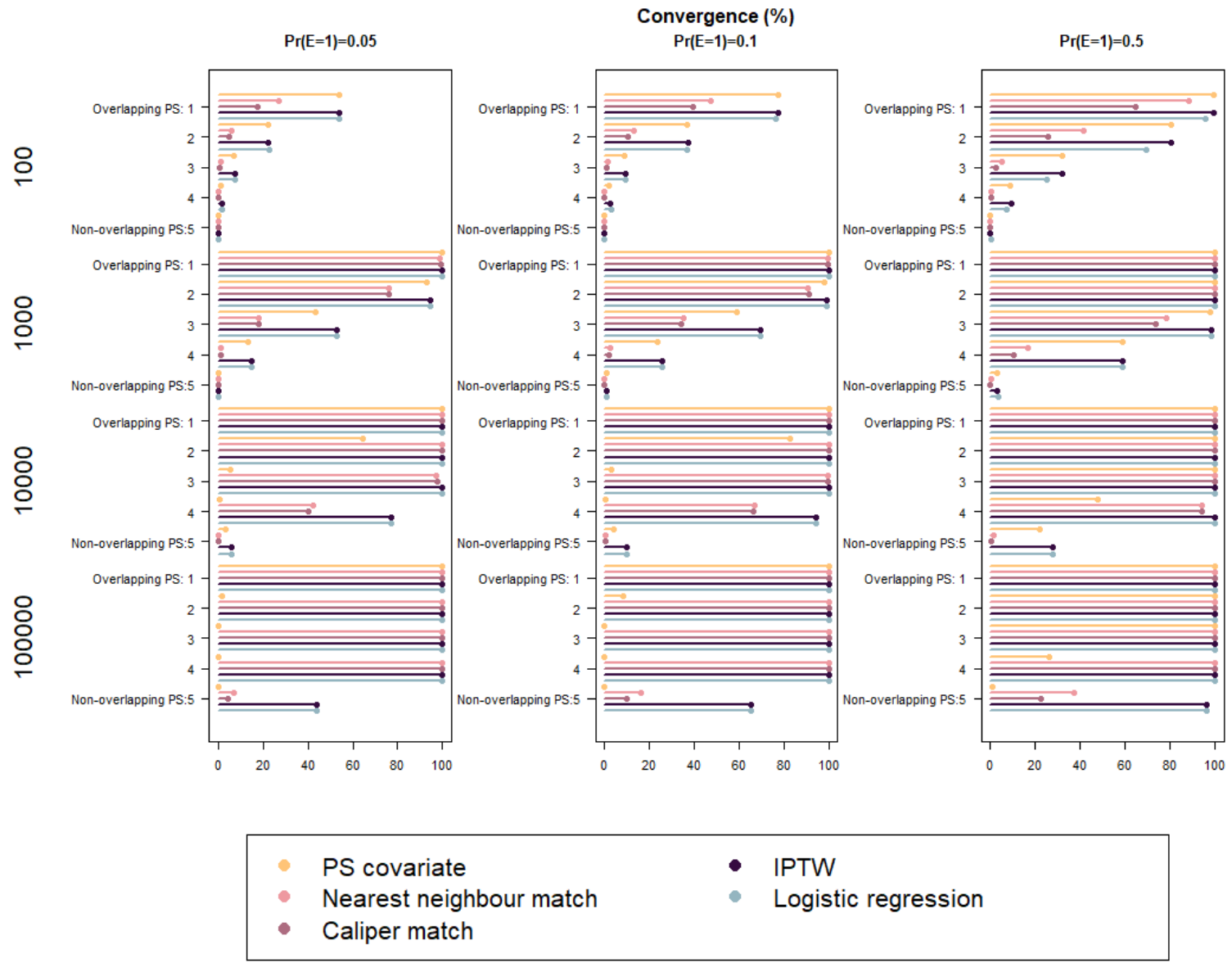

*S Figure 1: Convergence (%)*

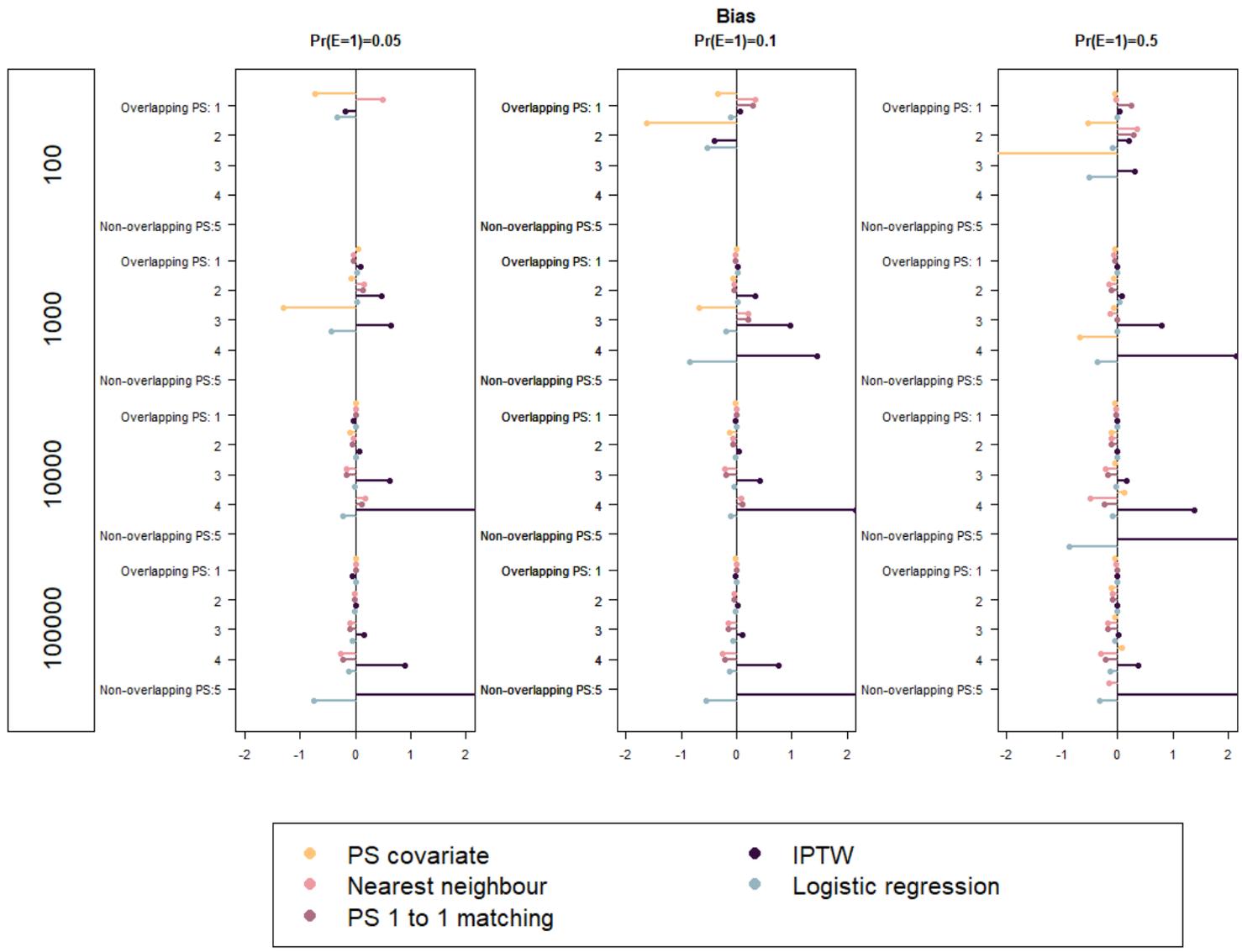

S Figure 2: Bias

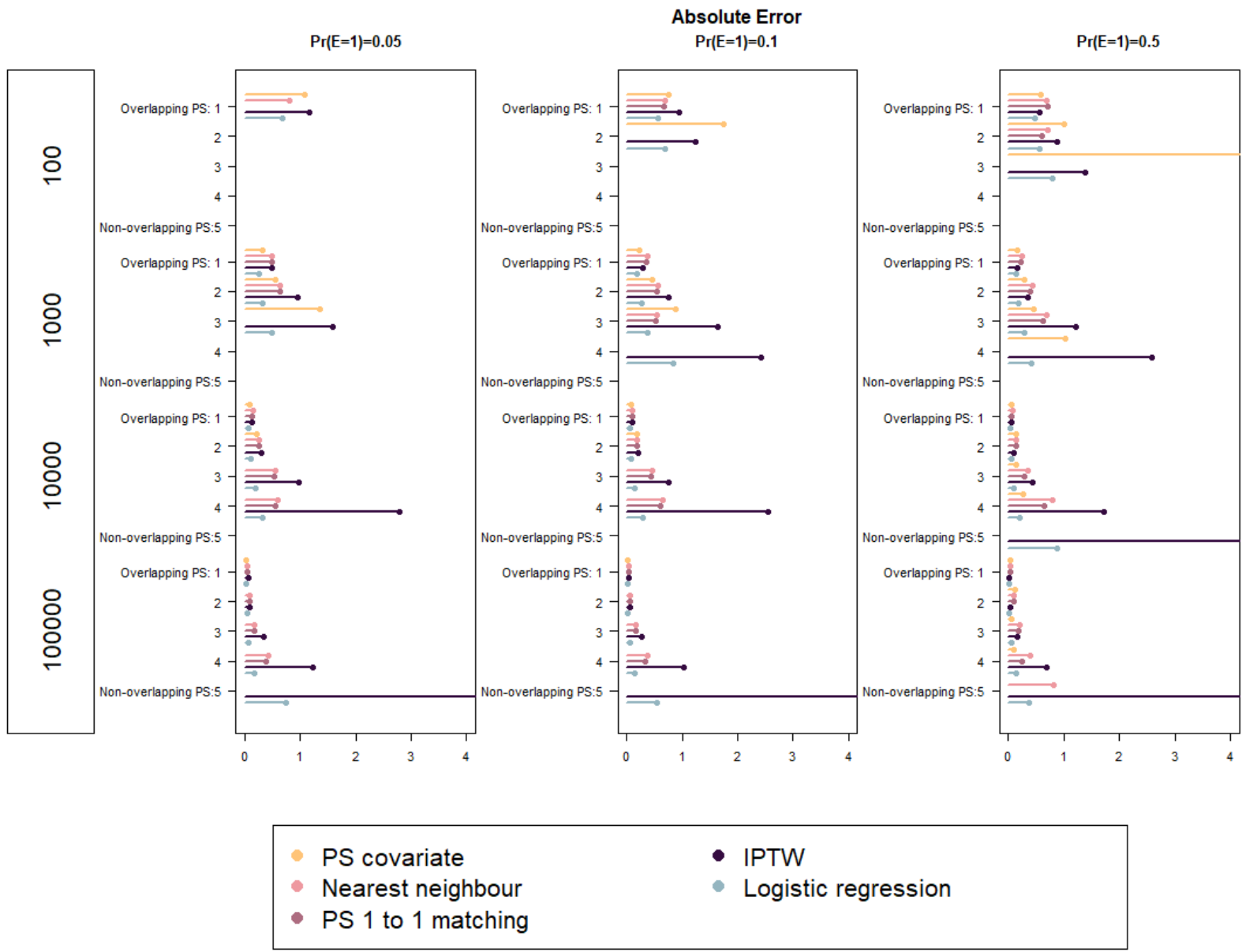

*S Figure 3: Absolute error*

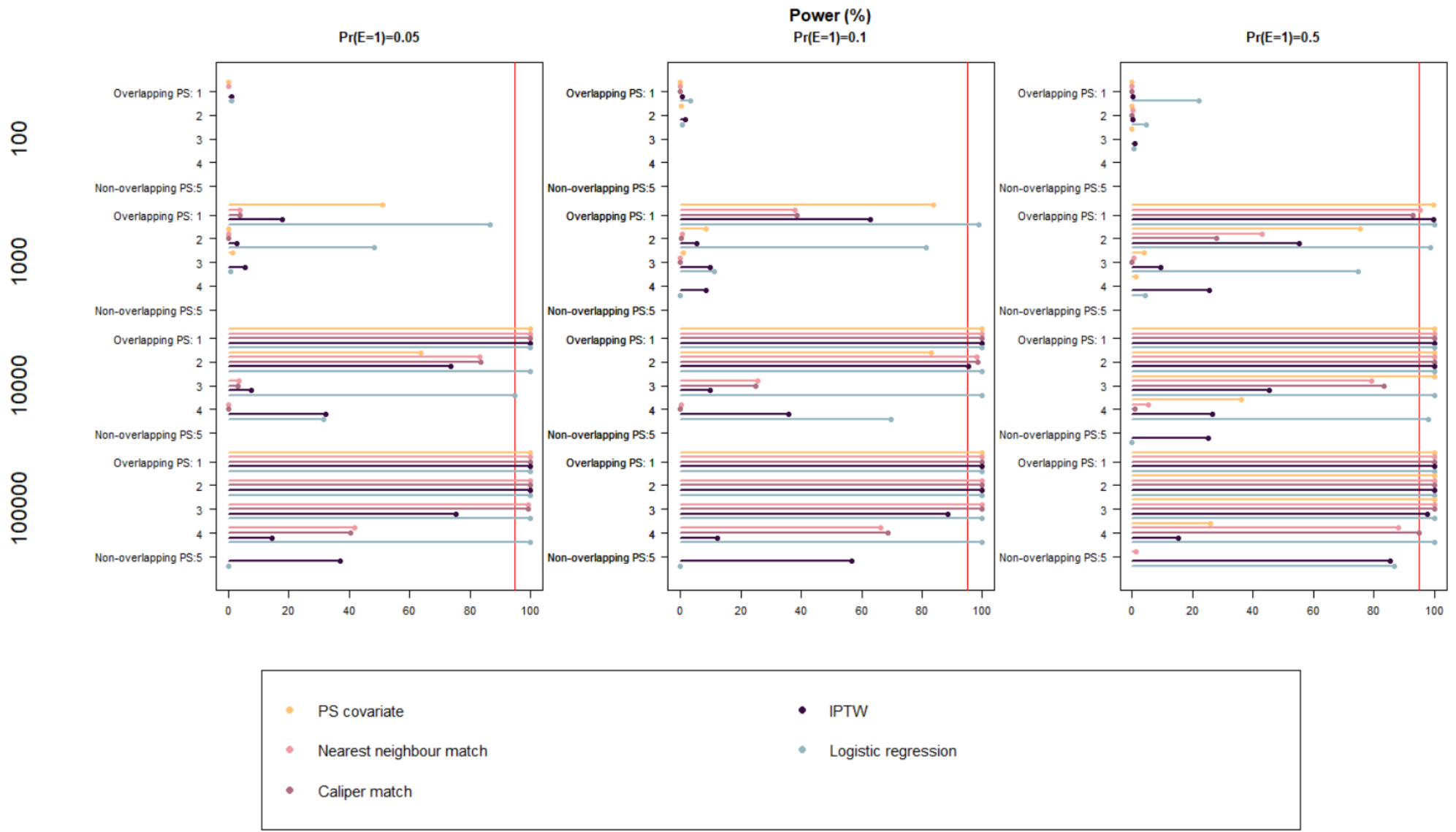

S Figure 4: Power (%)

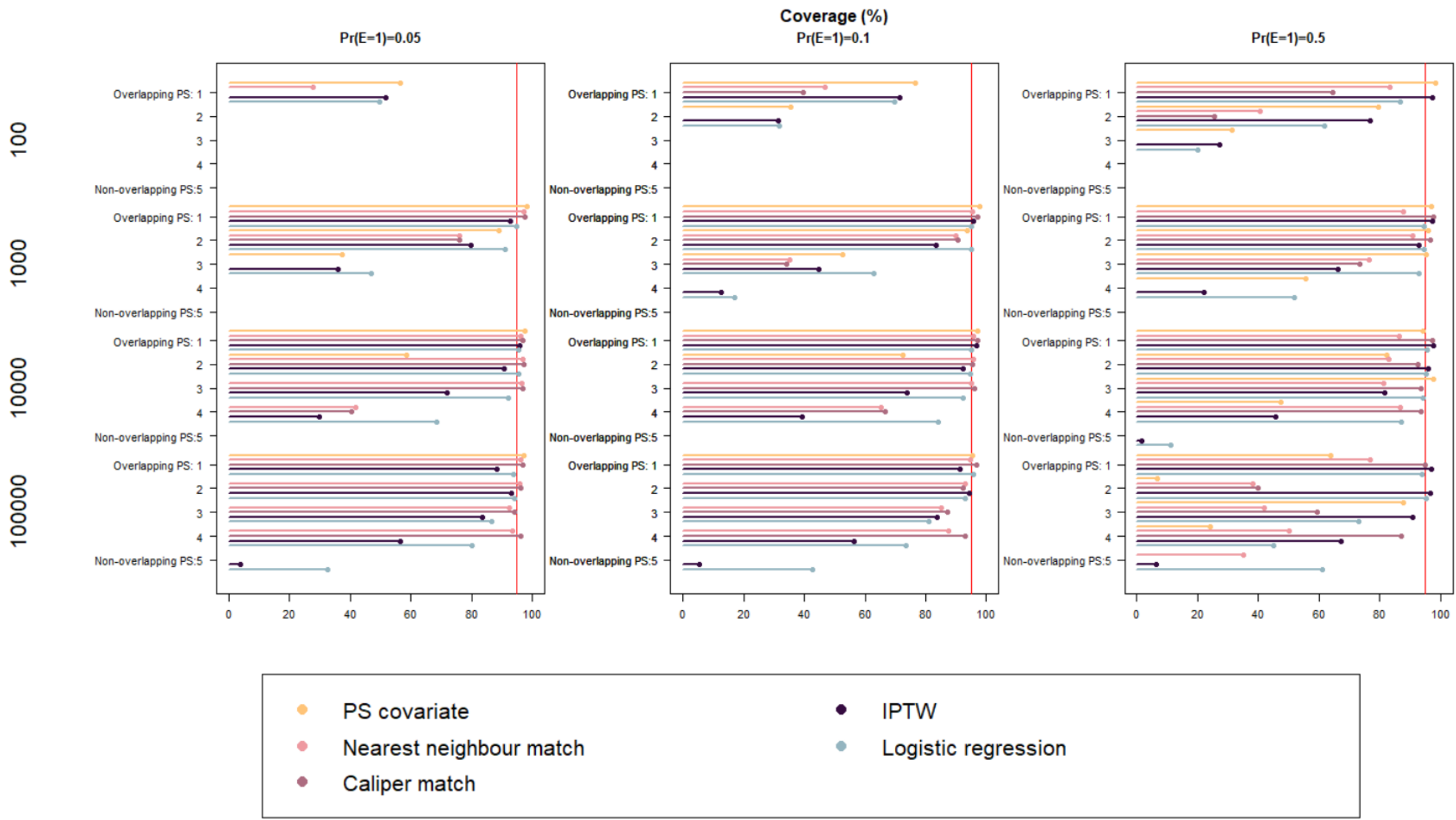

*S Figure 5: Coverage (%)*

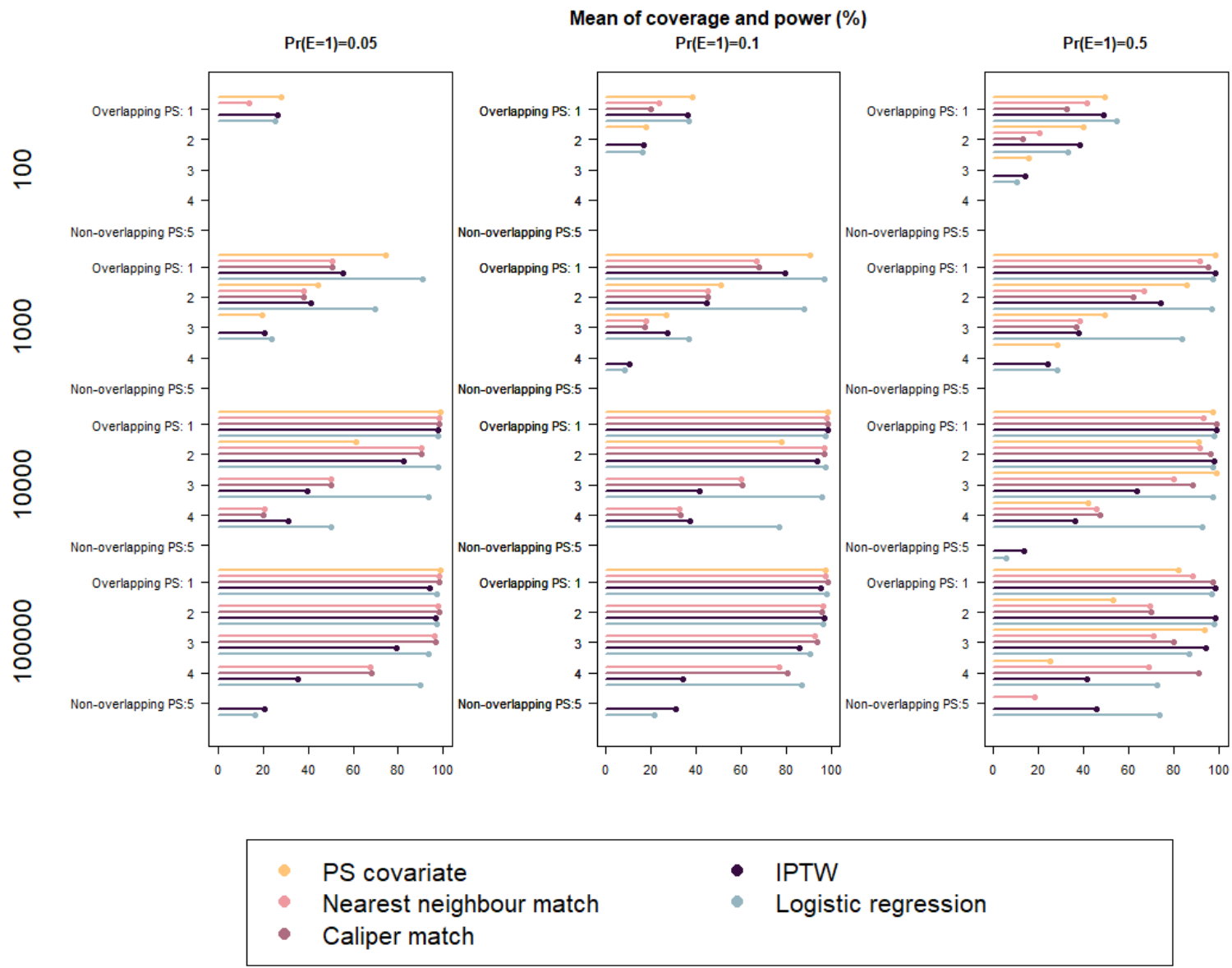

*S Figure 6: Mean of coverage and power (%)*

# Sensitivity 2: logit link outcome generation & unmeasured confounder

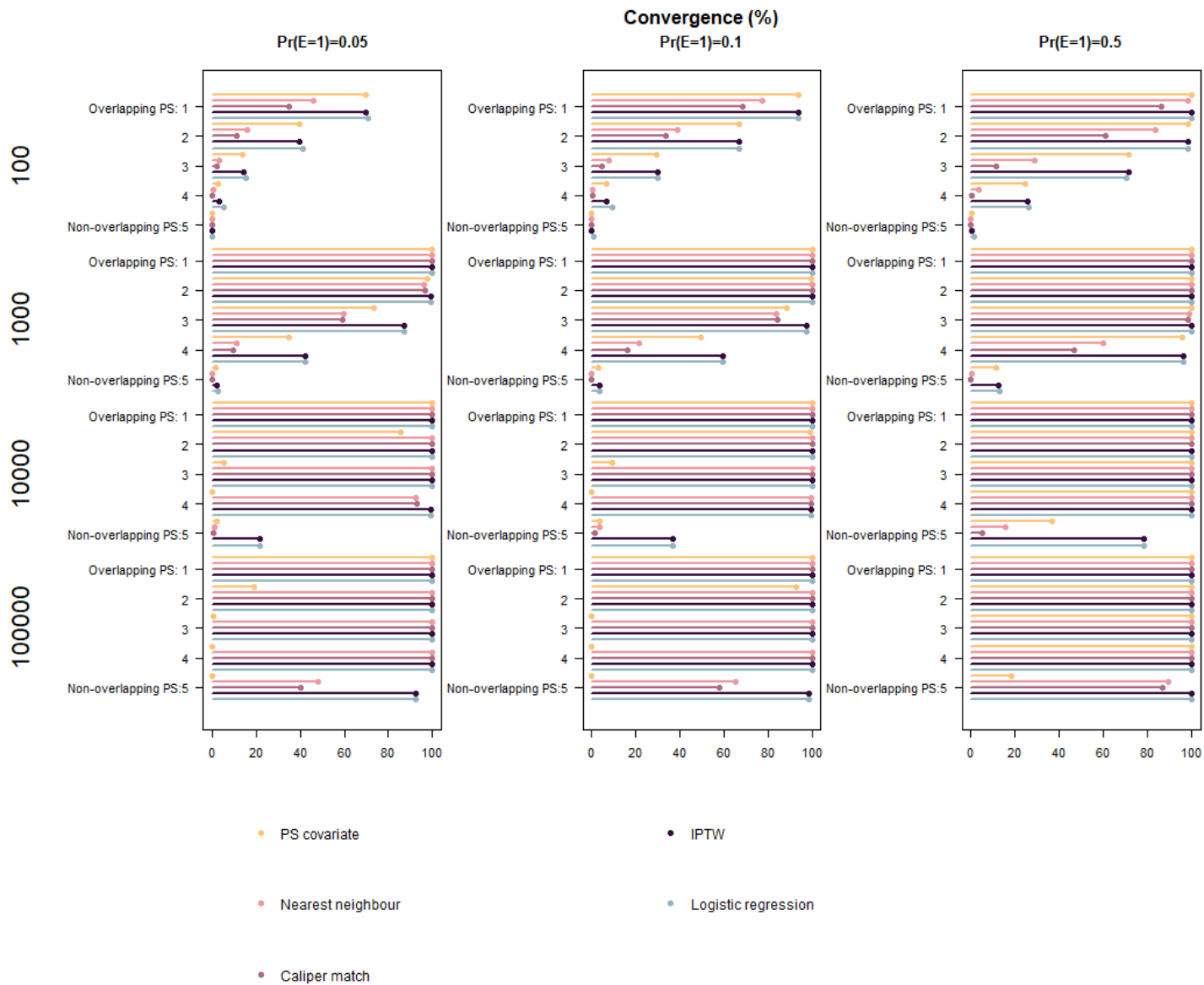

*S Figure 7: Convergence (%)*

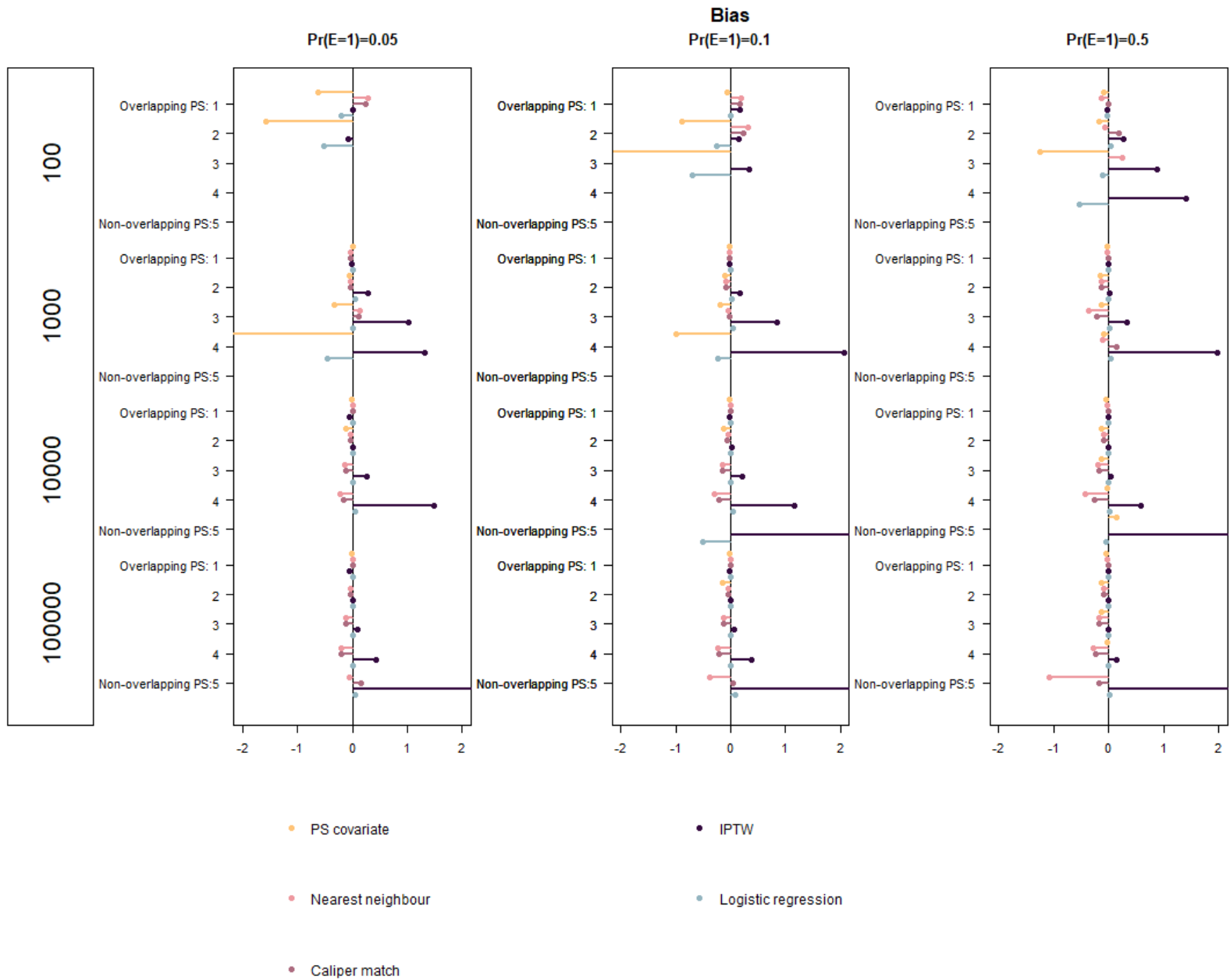

*S Figure 8: Bias*

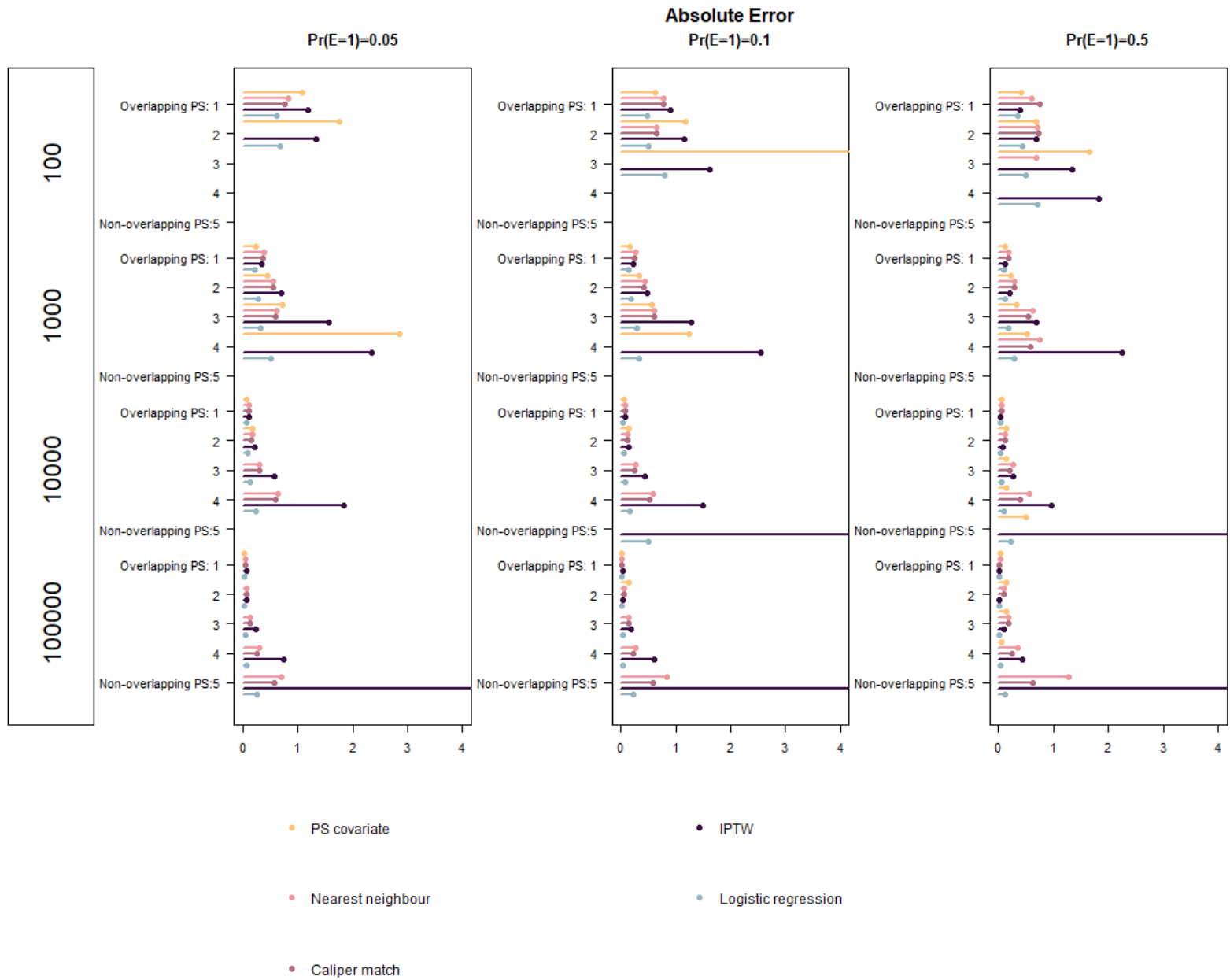

*S Figure 9: Absolute error*

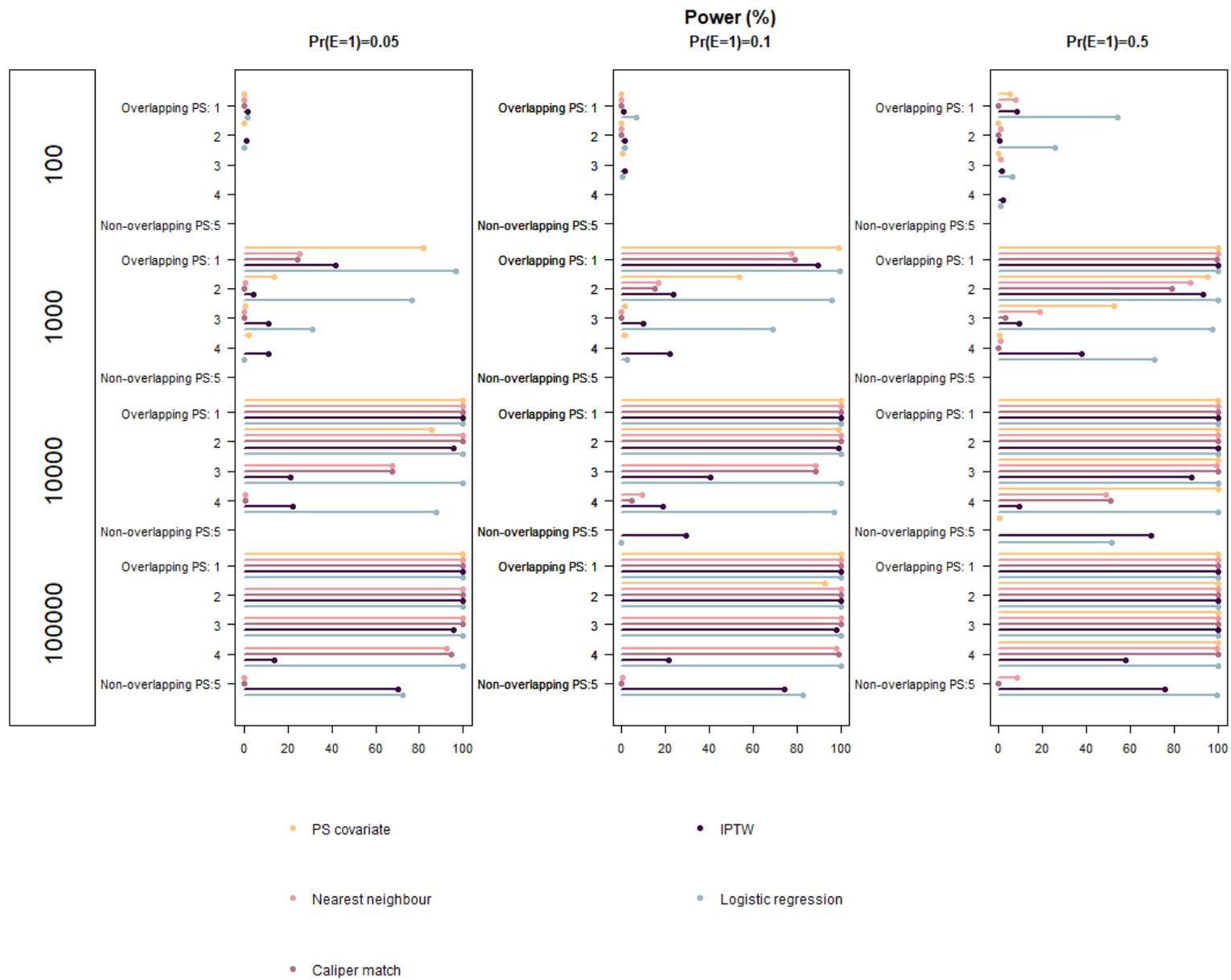

*S Figure 10: Power (%)*

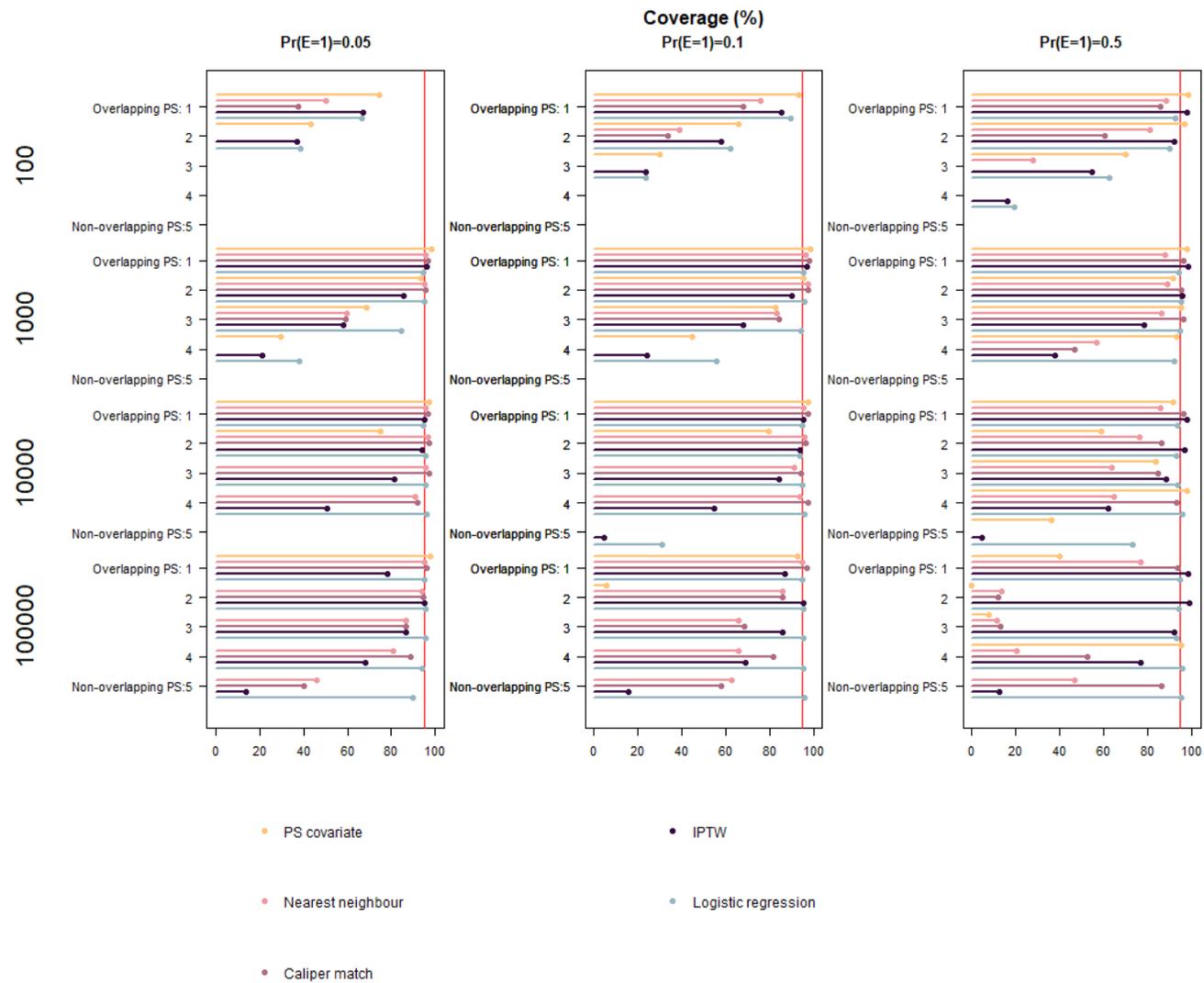

*S Figure 11: Coverage (%)*

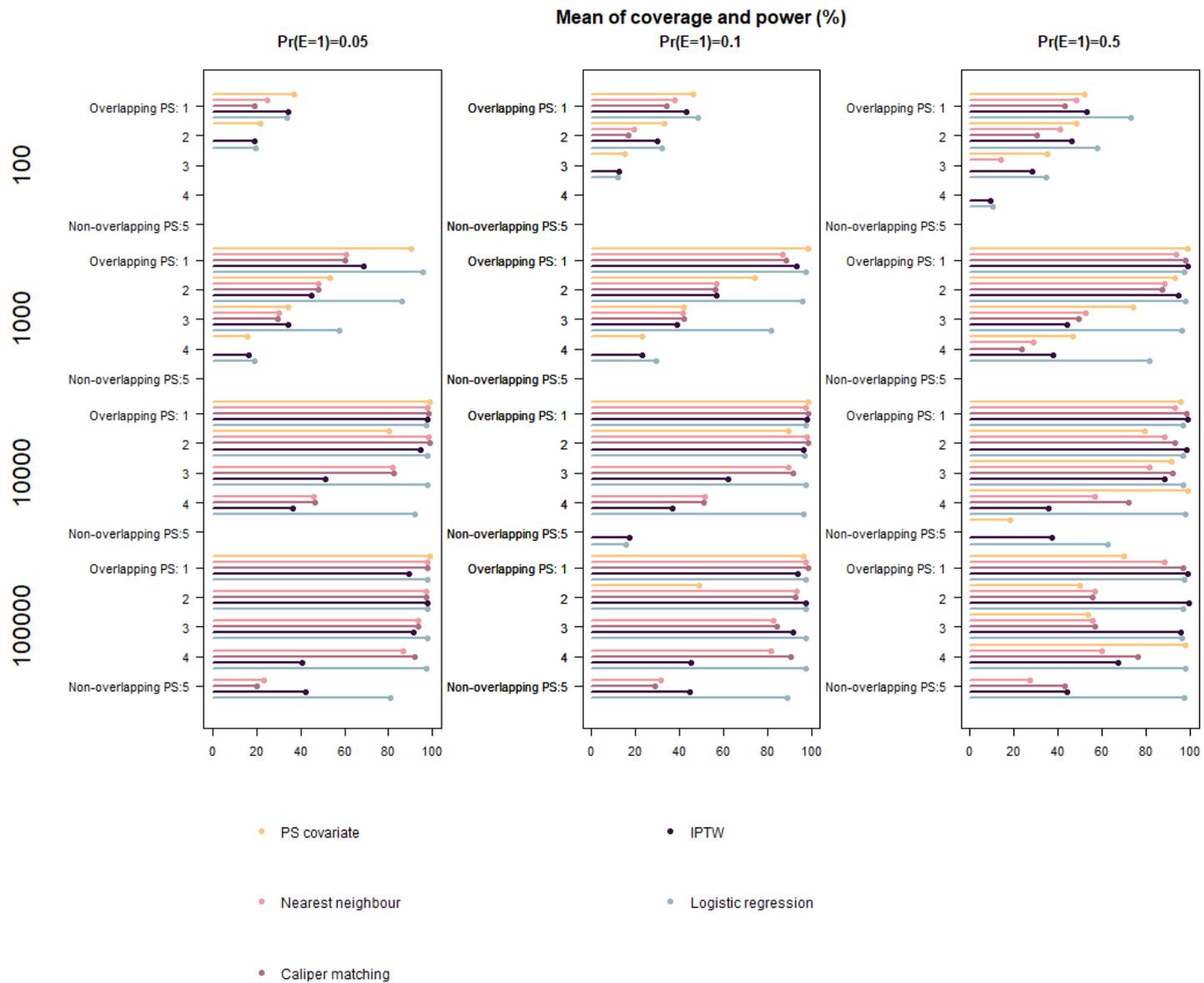

*S Figure 12: Mean of coverage and power (%)*

Sensitivity 3: complementary log-log link outcome generation & unmeasured confounder

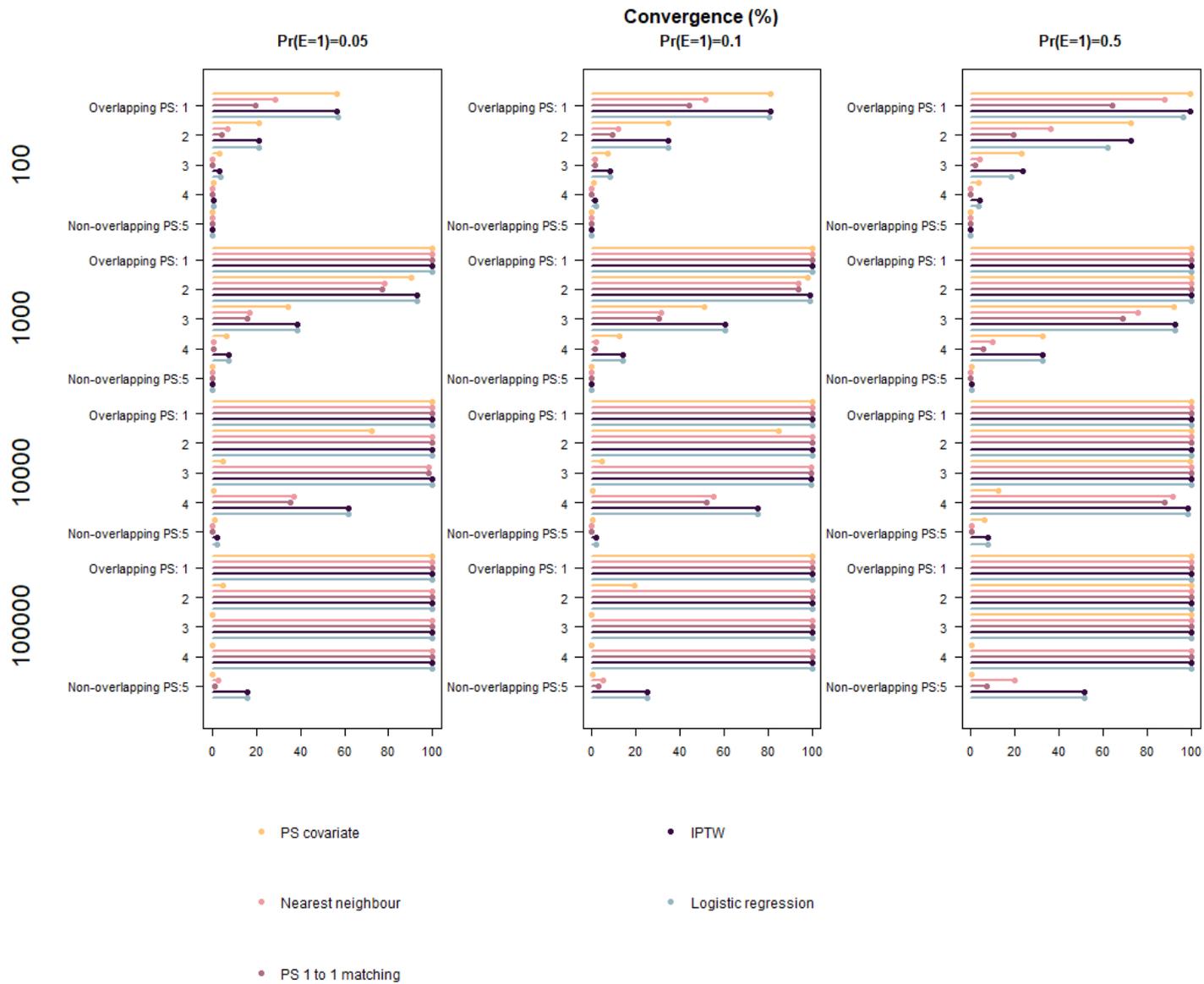

S Figure 13: Convergence (%)

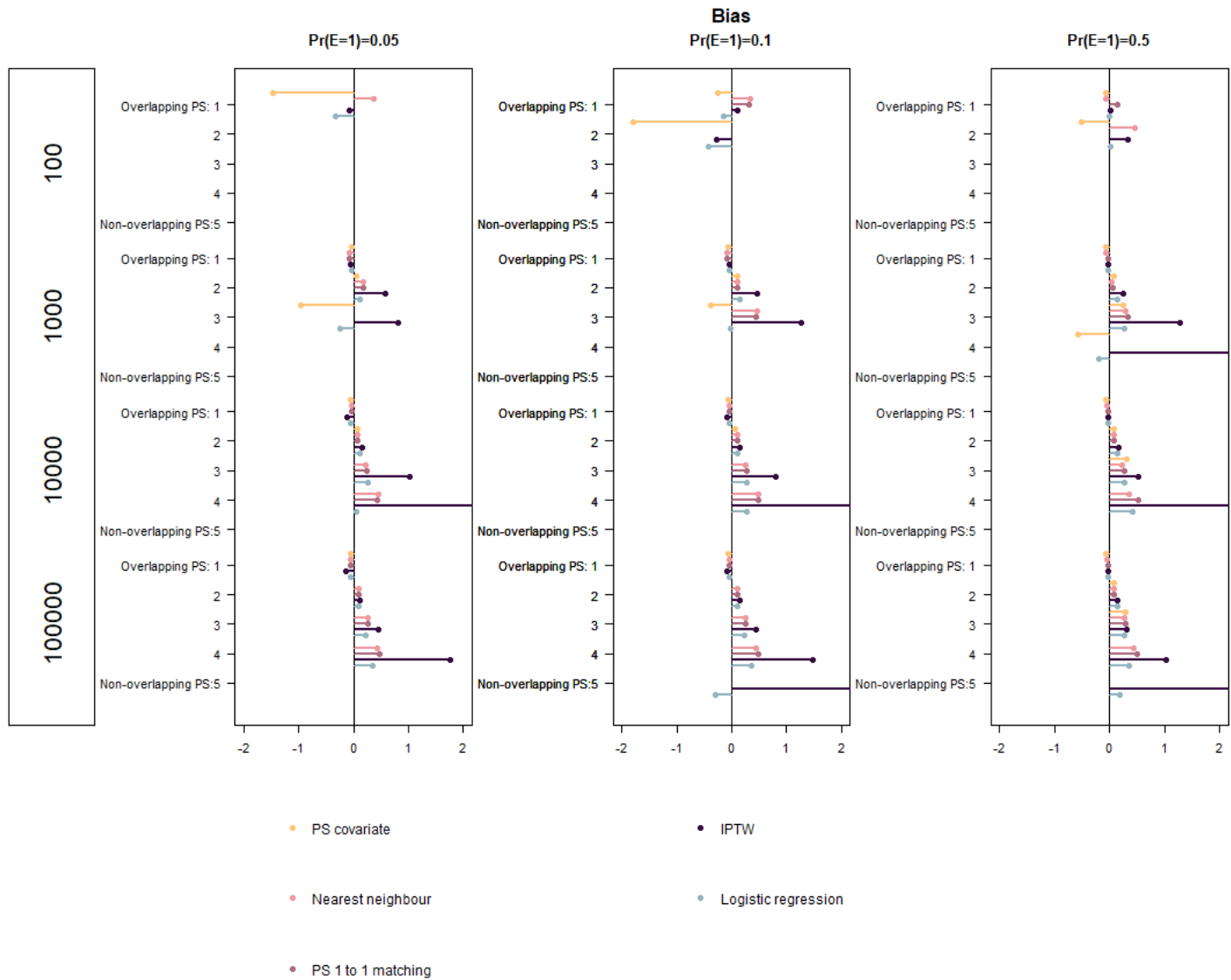

S Figure 14: Bias

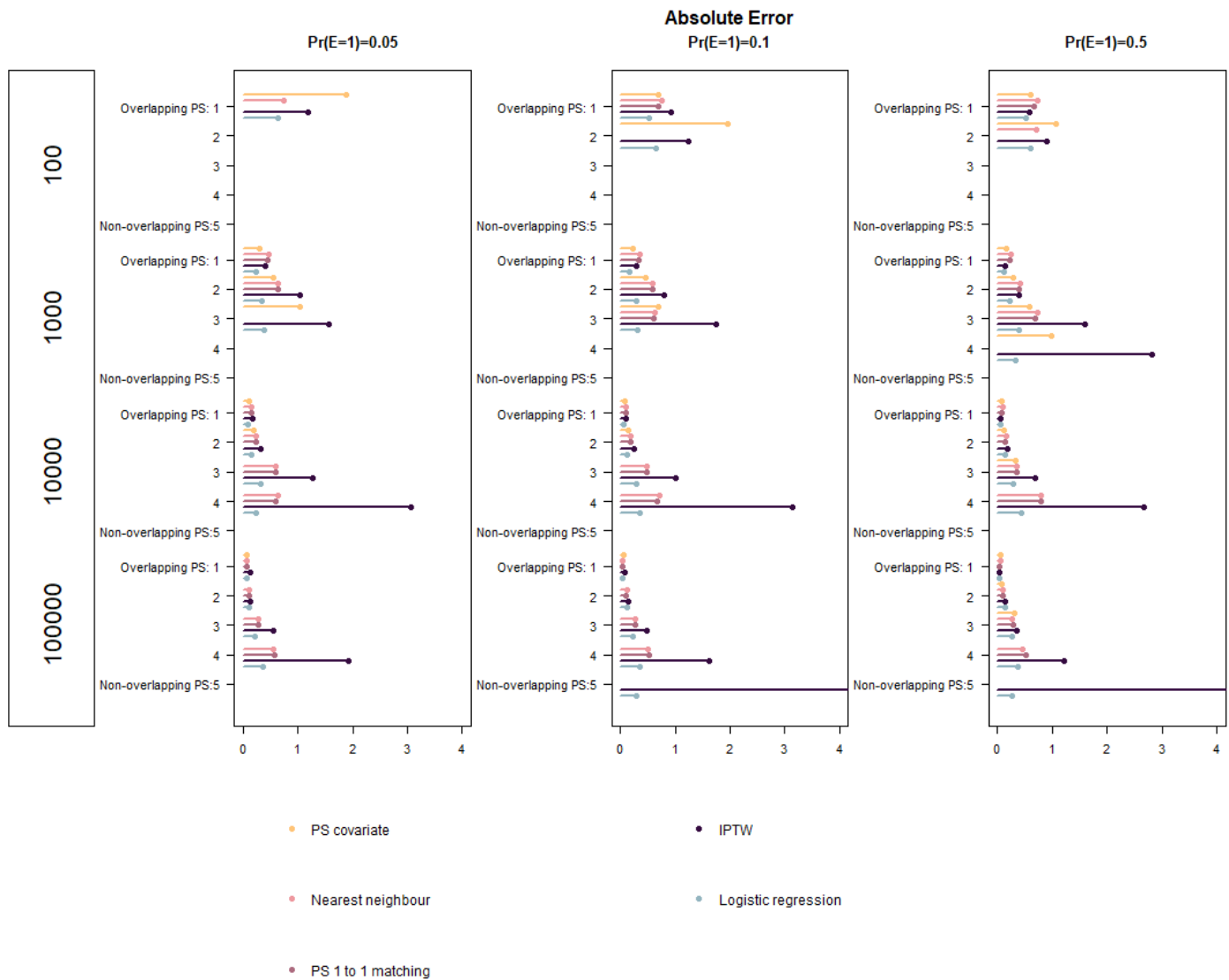

S Figure 15: Absolute error

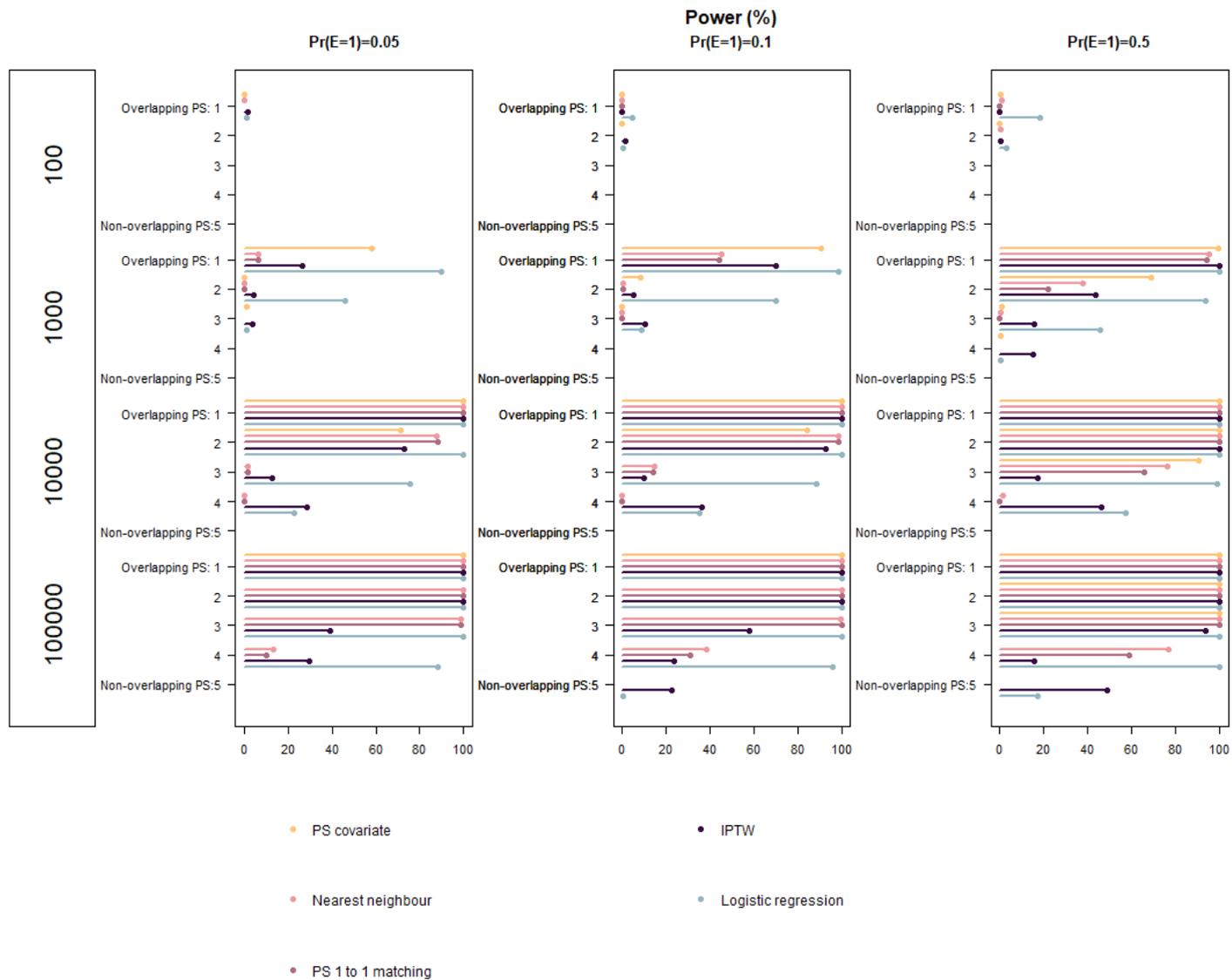

*S Figure 16: Power (%)*

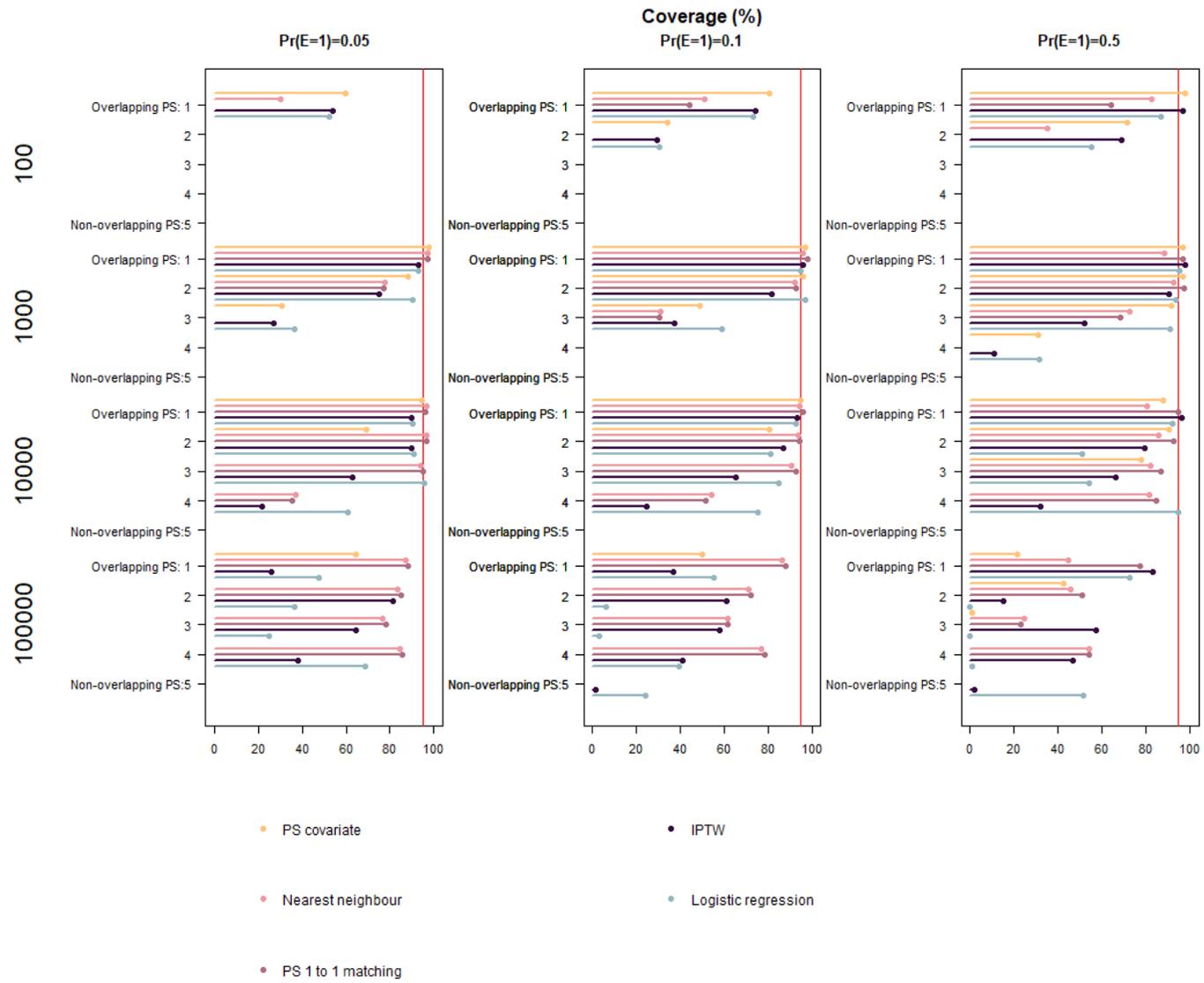

*S Figure 17: Coverage (%)*

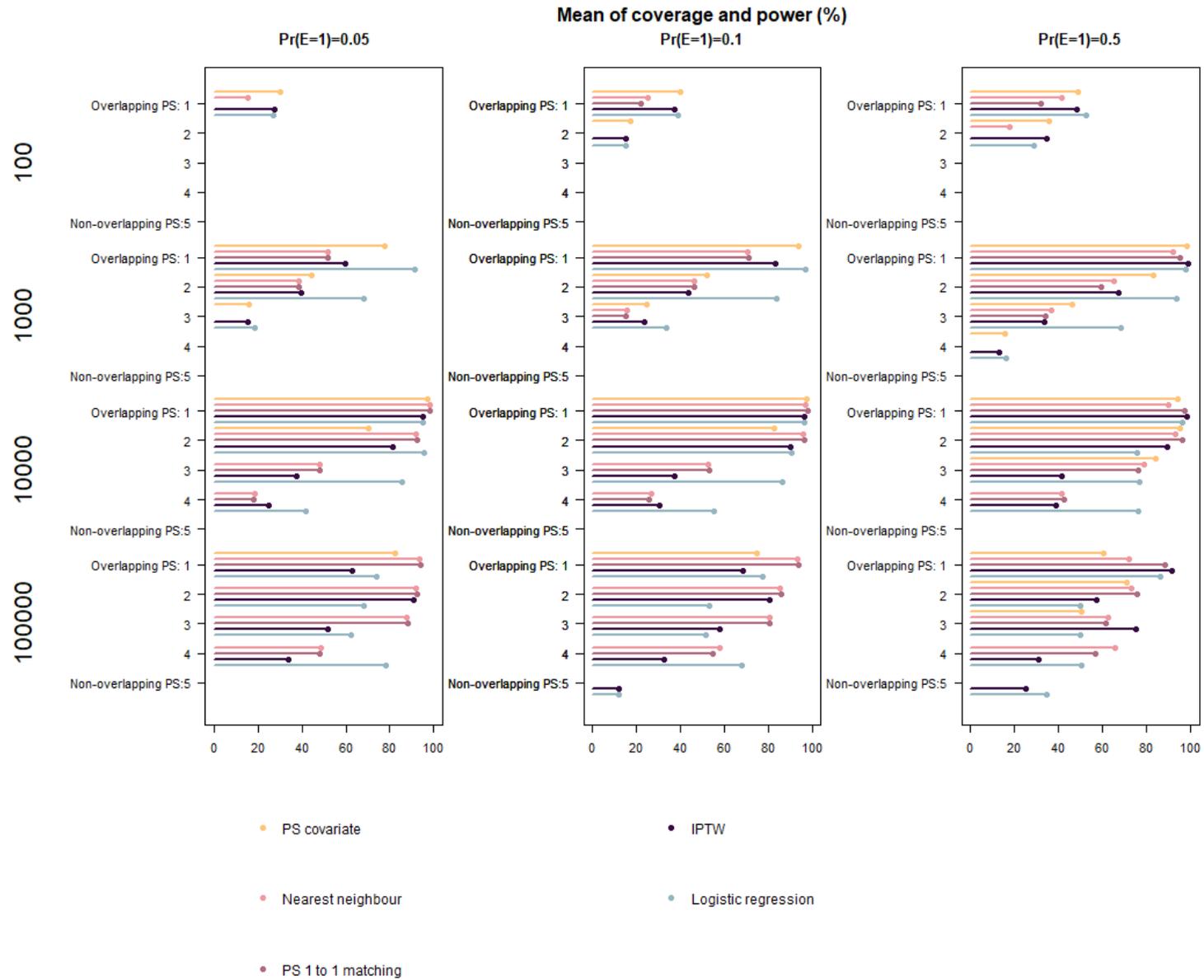

*S Figure 19: Mean of coverage and power (%)*

| n | Exposure probability | PS overlap scenario | Method | Convergence (%) | Bias | SE bias | Absolute error | SE absolute error | Power (%) | SE Power | Coverage (%) | SE coverage |
|---|---|---|---|---|---|---|---|---|---|---|---|---|
| 1.00E+05 | 0.05 | 1 | PS covariate | 100 | -0.006 | 2.98E-05 | 0.024288 | 1.82E-05 | 100 | 0 | 97.8 | 0 |
| 1.00E+05 | 0.05 | 2 | PS covariate | 13.6 | NA | NA | NA | NA | NA | NA | NA | NA |
| 1.00E+05 | 0.05 | 3 | PS covariate | 0 | NA | NA | NA | NA | NA | NA | NA | NA |
| 1.00E+05 | 0.05 | 4 | PS covariate | 0 | NA | NA | NA | NA | NA | NA | NA | NA |
| 1.00E+05 | 0.05 | 5 | PS covariate | 0 | NA | NA | NA | NA | NA | NA | NA | NA |
| 10000 | 0.05 | 1 | PS covariate | 100 | -0.00561 | 9.74E-05 | 0.07874 | 5.75E-05 | 100 | 0 | 98.2 | 0 |
| 10000 | 0.05 | 2 | PS covariate | 86.2 | -0.11707 | 0.000201 | 0.170456 | 0.00014 | 86.2 | 1.09067 | 78.7 | 1.09067 |
| 10000 | 0.05 | 3 | PS covariate | 5.3 | NA | NA | NA | NA | NA | NA | NA | NA |
| 10000 | 0.05 | 4 | PS covariate | 0.4 | NA | NA | NA | NA | NA | NA | NA | NA |
| 10000 | 0.05 | 5 | PS covariate | 2.4 | NA | NA | NA | NA | NA | NA | NA | NA |
| 1000 | 0.05 | 1 | PS covariate | 100 | 0.013638 | 0.000323 | 0.258058 | 0.000195 | 71.8 | 1.422941 | 98.8 | 1.422941 |
| 1000 | 0.05 | 2 | PS covariate | 98 | -0.06677 | 0.000586 | 0.44939 | 0.000372 | 8.3 | 0.872416 | 94.3 | 0.872416 |
| 1000 | 0.05 | 3 | PS covariate | 65.2 | -0.50877 | 0.001441 | 0.799317 | 0.001086 | 0.4 | 0.1996 | 60.9 | 0.1996 |
| 1000 | 0.05 | 4 | PS covariate | 27.6 | -2.71731 | 0.017789 | 2.792996 | 0.017634 | 2.3 | 0.474036 | 23.2 | 0.474036 |
| 1000 | 0.05 | 5 | PS covariate | 1.3 | NA | NA | NA | NA | NA | NA | NA | NA |
| 100 | 0.05 | 1 | PS covariate | 68.4 | -0.48205 | 0.002096 | 0.900096 | 0.001777 | 0 | 0 | 71.94474 | 0 |
| 100 | 0.05 | 2 | PS covariate | 37.1 | -3.72365 | 0.076944 | 3.856661 | 0.076896 | 0.111607 | 0.111545 | 40.625 | 0.111545 |
| 100 | 0.05 | 3 | PS covariate | 12.3 | NA | NA | NA | NA | NA | NA | NA | NA |
| 100 | 0.05 | 4 | PS covariate | 1.4 | NA | NA | NA | NA | NA | NA | NA | NA |
| 100 | 0.05 | 5 | PS covariate | 0 | NA | NA | NA | NA | NA | NA | NA | NA |
| 1.00E+05 | 0.05 | 1 | Nearest neighbour match | 100 | 0.005783 | 4.87E-05 | 0.039887 | 2.85E-05 | 100 | 0 | 96.7 | 0 |
| 1.00E+05 | 0.05 | 2 | Nearest neighbour match | 100 | -0.03161 | 6.89E-05 | 0.060404 | 4.57E-05 | 100 | 0 | 95.6 | 0 |
| 1.00E+05 | 0.05 | 3 | Nearest neighbour match | 100 | -0.11476 | 0.000128 | 0.140501 | 9.96E-05 | 100 | 0 | 88.1 | 0 |
| 1.00E+05 | 0.05 | 4 | Nearest neighbour match | 100 | -0.2207 | 0.0003 | 0.301781 | 0.000218 | 87 | 1.063485 | 87.7 | 1.063485 |
| 1.00E+05 | 0.05 | 5 | Nearest neighbour match | 34 | 0.15977 | 0.002549 | 0.649535 | 0.001749 | 0.4 | 0.1996 | 32.7 | 0.1996 |

**Supp Table 1a: Simulation results for exposure probability of 0.05 (continues in 1b)**

| n | Exposure probability | PS overlap scenario | Method | Convergence (%) | Bias | SE bias | Absolute error | SE absolute error | Power (%) | SE Power | Coverage (%) | SE coverage |
|---|---|---|---|---|---|---|---|---|---|---|---|---|
| 10000 | 0.05 | 1 | Nearest neighbour match | 100 | -0.00038 | 0.000157 | 0.125769 | 9.33E-05 | 100 | 0 | 96 | 0 |
| 10000 | 0.05 | 2 | Nearest neighbour match | 100 | -0.03023 | 0.000229 | 0.182409 | 0.000142 | 98.9 | 0.329833 | 95.8 | 0.329833 |
| 10000 | 0.05 | 3 | Nearest neighbour match | 100 | -0.14593 | 0.000464 | 0.365964 | 0.000321 | 47.2 | 1.578658 | 95.2 | 1.578658 |
| 10000 | 0.05 | 4 | Nearest neighbour match | 85 | -0.10145 | 0.000936 | 0.657924 | 0.00054 | 0.1 | 0.09995 | 83.9 | 0.09995 |
| 10000 | 0.05 | 5 | Nearest neighbour match | 0.9 | NA | NA | NA | NA | NA | NA | NA | NA |
| 1000 | 0.05 | 1 | Nearest neighbour match | 100 | -0.02621 | 0.000517 | 0.401647 | 0.000326 | 17.6 | 1.204259 | 96.9 | 1.204259 |
| 1000 | 0.05 | 2 | Nearest neighbour match | 94.3 | -0.09181 | 0.000758 | 0.573102 | 0.000463 | 0.3 | 0.172945 | 93.7 | 0.172945 |
| 1000 | 0.05 | 3 | Nearest neighbour match | 47.7 | 0.2392 | 0.001499 | 0.621684 | 0.000893 | 0 | 0 | 47.6 | 0 |
| 1000 | 0.05 | 4 | Nearest neighbour match | 8.8 | NA | NA | NA | NA | NA | NA | NA | NA |
| 1000 | 0.05 | 5 | Nearest neighbour match | 0 | NA | NA | NA | NA | NA | NA | NA | NA |
| 100 | 0.05 | 1 | Nearest neighbour match | 42.4 | 0.26166 | 0.002187 | 0.771822 | 0.001358 | 0 | 0 | 44.5271 | 0 |
| 100 | 0.05 | 2 | Nearest neighbour match | 14.7 | NA | NA | NA | NA | NA | NA | NA | NA |
| 100 | 0.05 | 3 | Nearest neighbour match | 1.9 | NA | NA | NA | NA | NA | NA | NA | NA |
| 100 | 0.05 | 4 | Nearest neighbour match | 0.3 | NA | NA | NA | NA | NA | NA | NA | NA |
| 100 | 0.05 | 5 | Nearest neighbour match | 0 | NA | NA | NA | NA | NA | NA | NA | NA |
| 1.00E+05 | 0.05 | 1 | Caliper match | 100 | 0.005765 | 4.77E-05 | 0.038881 | 2.82E-05 | 100 | 0 | 96.4 | 0 |
| 1.00E+05 | 0.05 | 2 | Caliper match | 100 | -0.03257 | 6.64E-05 | 0.058612 | 4.51E-05 | 100 | 0 | 96.1 | 0 |
| 1.00E+05 | 0.05 | 3 | Caliper match | 100 | -0.11779 | 0.000123 | 0.13966 | 9.69E-05 | 100 | 0 | 89.1 | 0 |
| 1.00E+05 | 0.05 | 4 | Caliper match | 100 | -0.20387 | 0.000282 | 0.277968 | 0.000209 | 87.2 | 1.056485 | 91 | 1.056485 |
| 1.00E+05 | 0.05 | 5 | Caliper match | 25.2 | 0.245413 | 0.0021 | 0.501002 | 0.001181 | 0 | 0 | 25.2 | 0 |
| 10000 | 0.05 | 1 | Caliper match | 100 | 0.000123 | 0.00015 | 0.119543 | 9.04E-05 | 100 | 0 | 96.8 | 0 |
| 10000 | 0.05 | 2 | Caliper match | 100 | -0.03578 | 0.000224 | 0.178132 | 0.00014 | 99 | 0.314643 | 96.3 | 0.314643 |
| 10000 | 0.05 | 3 | Caliper match | 100 | -0.14247 | 0.000448 | 0.358378 | 0.000304 | 43.9 | 1.569328 | 96.5 | 1.569328 |
| 10000 | 0.05 | 4 | Caliper match | 84.7 | -0.04771 | 0.000925 | 0.650842 | 0.000518 | 0 | 0 | 84.4 | 0 |
| 10000 | 0.05 | 5 | Caliper match | 0.4 | NA | NA | NA | NA | NA | NA | NA | NA |

**Supp Table 1b: Simulation results for exposure probability of 0.05 (continued from 1a, continues in 1c)**

| n | Exposure probability | PS overlap scenario | Method | Convergence (%) | Bias | SE bias | Absolute error | SE absolute error | Power (%) | SE Power | Coverage (%) | SE coverage |
|---|---|---|---|---|---|---|---|---|---|---|---|---|
| 1000 | 0.05 | 1 | Caliper match | 100 | -0.02414 | 0.000503 | 0.394209 | 0.000314 | 16.5 | 1.173776 | 97.6 | 1.173776 |
| 1000 | 0.05 | 2 | Caliper match | 94.4 | -0.07907 | 0.000751 | 0.56141 | 0.000465 | 0.2 | 0.14128 | 93.9 | 0.14128 |
| 1000 | 0.05 | 3 | Caliper match | 46.4 | 0.221823 | 0.001501 | 0.602537 | 0.00089 | 0 | 0 | 46.3 | 0 |
| 1000 | 0.05 | 4 | Caliper match | 7.1 | NA | NA | NA | NA | NA | NA | NA | NA |
| 1000 | 0.05 | 5 | Caliper match | 0 | NA | NA | NA | NA | NA | NA | NA | NA |
| 100 | 0.05 | 1 | Caliper match | 32 | 0.181802 | 0.002686 | 0.710504 | 0.00161 | 0 | 0 | 34.00637 | 0 |
| 100 | 0.05 | 2 | Caliper match | 10 | NA | NA | NA | NA | NA | NA | NA | NA |
| 100 | 0.05 | 3 | Caliper match | 1 | NA | NA | NA | NA | NA | NA | NA | NA |
| 100 | 0.05 | 4 | Caliper match | 0.1 | NA | NA | NA | NA | NA | NA | NA | NA |
| 100 | 0.05 | 5 | Caliper match | 0 | NA | NA | NA | NA | NA | NA | NA | NA |
| 1.00E+05 | 0.05 | 1 | IPTW | 100 | -0.05477 | 4.15E-05 | 0.05839 | 3.62E-05 | 100 | 0 | 81.3 | 0 |
| 1.00E+05 | 0.05 | 2 | IPTW | 100 | 0.009911 | 9.22E-05 | 0.074407 | 5.54E-05 | 100 | 0 | 93.8 | 0 |
| 1.00E+05 | 0.05 | 3 | IPTW | 100 | 0.082119 | 0.000305 | 0.249743 | 0.000193 | 93 | 0.806846 | 85.7 | 0.806846 |
| 1.00E+05 | 0.05 | 4 | IPTW | 100 | 0.529221 | 0.000869 | 0.834237 | 0.000582 | 9.4 | 0.922843 | 65.3 | 0.922843 |
| 1.00E+05 | 0.05 | 5 | IPTW | 82.9 | 5.602679 | 0.003796 | 5.815961 | 0.003296 | 66.8 | 1.489215 | 9.8 | 1.489215 |
| 10000 | 0.05 | 1 | IPTW | 100 | -0.04564 | 0.000137 | 0.114606 | 8.82E-05 | 100 | 0 | 95.4 | 0 |
| 10000 | 0.05 | 2 | IPTW | 100 | 0.043459 | 0.000287 | 0.229254 | 0.000177 | 92.1 | 0.852989 | 91.6 | 0.852989 |
| 10000 | 0.05 | 3 | IPTW | 100 | 0.342119 | 0.000798 | 0.684793 | 0.000534 | 11.7 | 1.01642 | 77.7 | 1.01642 |
| 10000 | 0.05 | 4 | IPTW | 98.2 | 1.779791 | 0.001975 | 2.121554 | 0.001586 | 29 | 1.434922 | 44 | 1.434922 |
| 10000 | 0.05 | 5 | IPTW | 17.2 | NA | NA | NA | NA | NA | NA | NA | NA |
| 1000 | 0.05 | 1 | IPTW | 100 | 0.009664 | 0.000488 | 0.382063 | 0.000303 | 35.8 | 1.516034 | 94.3 | 1.516034 |
| 1000 | 0.05 | 2 | IPTW | 99 | 0.326597 | 0.000953 | 0.740059 | 0.000676 | 4 | 0.619677 | 85.5 | 0.619677 |
| 1000 | 0.05 | 3 | IPTW | 76.6 | 0.944739 | 0.002334 | 1.606316 | 0.001603 | 9.1 | 0.9095 | 52.1 | 0.9095 |
| 1000 | 0.05 | 4 | IPTW | 32.5 | 1.239097 | 0.007387 | 2.17999 | 0.0049 | 8.3 | 0.872416 | 17.8 | 0.872416 |
| 1000 | 0.05 | 5 | IPTW | 1.5 | NA | NA | NA | NA | NA | NA | NA | NA |

**Supp Table 1c: Simulation results for exposure probability of 0.05 (continued from 1b, continues in 1d)**

| n | Exposure probability | PS overlap scenario | Method | Convergence (%) | Bias | SE bias | Absolute error | SE absolute error | Power (%) | SE Power | Coverage (%) | SE coverage |
|---|---|---|---|---|---|---|---|---|---|---|---|---|
| 100 | 0.05 | 1 | IPTW | 68.5 | -0.09842 | 0.002126 | 1.16657 | 0.001279 | 0.850159 | 0.299296 | 64.93092 | 0.299296 |
| 100 | 0.05 | 2 | IPTW | 37.1 | -0.23603 | 0.00454 | 1.347151 | 0.002792 | 1.004464 | 0.333136 | 34.15179 | 0.333136 |
| 100 | 0.05 | 3 | IPTW | 12.7 | NA | NA | NA | NA | NA | NA | NA | NA |
| 100 | 0.05 | 4 | IPTW | 1.5 | NA | NA | NA | NA | NA | NA | NA | NA |
| 100 | 0.05 | 5 | IPTW | 0 | NA | NA | NA | NA | NA | NA | NA | NA |
| 1.00E+05 | 0.05 | 1 | Logistic regression | 100 | 0.000259 | 2.58E-05 | 0.02022 | 1.61E-05 | 100 | 0 | 95.6 | 0 |
| 1.00E+05 | 0.05 | 2 | Logistic regression | 100 | 0.000473 | 3.35E-05 | 0.02669 | 2.03E-05 | 100 | 0 | 95.7 | 0 |
| 1.00E+05 | 0.05 | 3 | Logistic regression | 100 | 0.000378 | 5.11E-05 | 0.040484 | 3.11E-05 | 100 | 0 | 96.5 | 0 |
| 1.00E+05 | 0.05 | 4 | Logistic regression | 100 | 0.007157 | 0.0001 | 0.07995 | 6.08E-05 | 100 | 0 | 94.3 | 0 |
| 1.00E+05 | 0.05 | 5 | Logistic regression | 82.9 | -0.01003 | 0.000372 | 0.26434 | 0.000192 | 50.6 | 1.581025 | 78.6 | 1.581025 |
| 10000 | 0.05 | 1 | Logistic regression | 100 | 0.002097 | 8.49E-05 | 0.068273 | 5.04E-05 | 100 | 0 | 94.7 | 0 |
| 10000 | 0.05 | 2 | Logistic regression | 100 | 0.003162 | 0.000108 | 0.084887 | 6.65E-05 | 100 | 0 | 95 | 0 |
| 10000 | 0.05 | 3 | Logistic regression | 100 | 0.018052 | 0.000179 | 0.136215 | 0.000117 | 99.2 | 0.281709 | 94.8 | 0.281709 |
| 10000 | 0.05 | 4 | Logistic regression | 98.2 | 0.071266 | 0.000343 | 0.26686 | 0.000222 | 77 | 1.330789 | 94.9 | 1.330789 |
| 10000 | 0.05 | 5 | Logistic regression | 17.2 | NA | NA | NA | NA | NA | NA | NA | NA |
| 1000 | 0.05 | 1 | Logistic regression | 100 | 0.012748 | 0.000271 | 0.215192 | 0.000166 | 94.8 | 0.702111 | 95.5 | 0.702111 |
| 1000 | 0.05 | 2 | Logistic regression | 99 | 0.038222 | 0.00037 | 0.282818 | 0.000239 | 74.9 | 1.371127 | 95.2 | 1.371127 |
| 1000 | 0.05 | 3 | Logistic regression | 76.6 | -0.08687 | 0.000537 | 0.337659 | 0.000326 | 14.4 | 1.110243 | 72.7 | 1.110243 |
| 1000 | 0.05 | 4 | Logistic regression | 32.5 | -0.59497 | 0.001088 | 0.608218 | 0.001016 | 0 | 0 | 27.9 | 0 |
| 1000 | 0.05 | 5 | Logistic regression | 1.7 | NA | NA | NA | NA | NA | NA | NA | NA |
| 100 | 0.05 | 1 | Logistic regression | 69.2 | -0.22644 | 0.00103 | 0.596547 | 0.00065 | 1.6 | 0.396787 | 65.2 | 0.396787 |
| 100 | 0.05 | 2 | Logistic regression | 38.1 | -0.602 | 0.001945 | 0.747228 | 0.001559 | 0.1 | 0.09995 | 33.7 | 0.09995 |
| 100 | 0.05 | 3 | Logistic regression | 14.1 | NA | NA | NA | NA | NA | NA | NA | NA |
| 100 | 0.05 | 4 | Logistic regression | 3 | NA | NA | NA | NA | NA | NA | NA | NA |
| 100 | 0.05 | 5 | Logistic regression | 0 | NA | NA | NA | NA | NA | NA | NA | NA |

**Supp Table 1d: Simulation results for exposure probability of 0.05 (continued from 1c)**

| n | Exposure probability | PS overlap scenario | Method | Convergence (%) | Bias | SE bias | Absolute error | SE absolute error | Power (%) | SE Power | Coverage (%) | SE coverage |
|---|---|---|---|---|---|---|---|---|---|---|---|---|
| 1.00E+05 | 0.1 | 1 | PS covariate | 100 | -0.01704 | 2.39E-05 | 0.023464 | 1.76E-05 | 100 | 0 | 92.7 | 0 |
| 1.00E+05 | 0.1 | 2 | PS covariate | 69.2 | -0.12569 | 6.13E-05 | 0.125731 | 6.11E-05 | 69.2 | 1.459918 | 11 | 1.459918 |
| 1.00E+05 | 0.1 | 3 | PS covariate | 0 | NA | NA | NA | NA | NA | NA | NA | NA |
| 1.00E+05 | 0.1 | 4 | PS covariate | 0 | NA | NA | NA | NA | NA | NA | NA | NA |
| 1.00E+05 | 0.1 | 5 | PS covariate | 0 | NA | NA | NA | NA | NA | NA | NA | NA |
| 10000 | 0.1 | 1 | PS covariate | 100 | -0.02084 | 7.10E-05 | 0.059675 | 4.37E-05 | 100 | 0 | 98.1 | 0 |
| 10000 | 0.1 | 2 | PS covariate | 96.4 | -0.12237 | 0.000136 | 0.147492 | 0.000106 | 96.4 | 0.589101 | 81.5 | 0.589101 |
| 10000 | 0.1 | 3 | PS covariate | 5 | NA | NA | NA | NA | NA | NA | NA | NA |
| 10000 | 0.1 | 4 | PS covariate | 0.1 | NA | NA | NA | NA | NA | NA | NA | NA |
| 10000 | 0.1 | 5 | PS covariate | 3.4 | NA | NA | NA | NA | NA | NA | NA | NA |
| 1000 | 0.1 | 1 | PS covariate | 100 | 0.002621 | 0.000247 | 0.19397 | 0.000153 | 96.2 | 0.604616 | 97.6 | 0.604616 |
| 1000 | 0.1 | 2 | PS covariate | 99.8 | -0.11358 | 0.000436 | 0.359156 | 0.000271 | 39 | 1.542401 | 94.5 | 1.542401 |
| 1000 | 0.1 | 3 | PS covariate | 83.3 | -0.21604 | 0.000931 | 0.63722 | 0.00059 | 0.6 | 0.244213 | 78.5 | 0.244213 |
| 1000 | 0.1 | 4 | PS covariate | 45.4 | -1.31846 | 0.004582 | 1.489077 | 0.004321 | 2.3 | 0.474036 | 39.9 | 0.474036 |
| 1000 | 0.1 | 5 | PS covariate | 2.2 | NA | NA | NA | NA | NA | NA | NA | NA |
| 100 | 0.1 | 1 | PS covariate | 89.4 | -0.11205 | 0.000941 | 0.648192 | 0.000612 | 0 | 0 | 88.4 | 0 |
| 100 | 0.1 | 2 | PS covariate | 57.8 | -1.15622 | 0.004382 | 1.362167 | 0.004201 | 0.501505 | 0.223717 | 55.4664 | 0.223717 |
| 100 | 0.1 | 3 | PS covariate | 21.9 | NA | NA | NA | NA | NA | NA | NA | NA |
| 100 | 0.1 | 4 | PS covariate | 4.8 | NA | NA | NA | NA | NA | NA | NA | NA |
| 100 | 0.1 | 5 | PS covariate | 0 | NA | NA | NA | NA | NA | NA | NA | NA |
| 1.00E+05 | 0.1 | 1 | Nearest neighbour match | 100 | 0.002547 | 3.70E-05 | 0.029486 | 2.25E-05 | 100 | 0 | 95.1 | 0 |
| 1.00E+05 | 0.1 | 2 | Nearest neighbour match | 100 | -0.04933 | 5.48E-05 | 0.06059 | 4.20E-05 | 100 | 0 | 86.7 | 0 |
| 1.00E+05 | 0.1 | 3 | Nearest neighbour match | 100 | -0.13985 | 0.000105 | 0.149107 | 9.17E-05 | 100 | 0 | 70.5 | 0 |
| 1.00E+05 | 0.1 | 4 | Nearest neighbour match | 100 | -0.2244 | 0.000258 | 0.281358 | 0.000194 | 95.3 | 0.669261 | 75.8 | 0.669261 |
| 1.00E+05 | 0.1 | 5 | Nearest neighbour match | 53.4 | -0.18803 | 0.001595 | 0.707596 | 0.000954 | 0.3 | 0.172945 | 51.6 | 0.172945 |

**Supp table 2a: Simulation results for exposure probability of 0.1 (continues in 2b)**

| n | Exposure probability | PS overlap scenario | Method | Convergence (%) | Bias | SE bias | Absolute error | SE absolute error | Power (%) | SE Power | Coverage (%) | SE coverage |
|---|---|---|---|---|---|---|---|---|---|---|---|---|
| 10000 | 0.1 | 1 | Nearest neighbour match | 100 | -0.001 | 0.000106 | 0.085809 | 6.30E-05 | 100 | 0 | 96.8 | 0 |
| 10000 | 0.1 | 2 | Nearest neighbour match | 100 | -0.04348 | 0.000175 | 0.142922 | 0.000109 | 99.9 | 0.09995 | 95.5 | 0.09995 |
| 10000 | 0.1 | 3 | Nearest neighbour match | 100 | -0.15187 | 0.000339 | 0.292323 | 0.000229 | 77.8 | 1.314215 | 93.9 | 1.314215 |
| 10000 | 0.1 | 4 | Nearest neighbour match | 95.4 | -0.32307 | 0.000758 | 0.625226 | 0.000509 | 2.2 | 0.463853 | 92.9 | 0.463853 |
| 10000 | 0.1 | 5 | Nearest neighbour match | 2.3 | NA | NA | NA | NA | NA | NA | NA | NA |
| 1000 | 0.1 | 1 | Nearest neighbour match | 100 | -0.00278 | 0.000374 | 0.295212 | 0.00023 | 68.4 | 1.470184 | 95.5 | 1.470184 |
| 1000 | 0.1 | 2 | Nearest neighbour match | 99.4 | -0.11419 | 0.000613 | 0.472871 | 0.000403 | 8.2 | 0.867617 | 97.1 | 0.867617 |
| 1000 | 0.1 | 3 | Nearest neighbour match | 72.6 | 0.091478 | 0.001102 | 0.659531 | 0.000635 | 0 | 0 | 71.8 | 0 |
| 1000 | 0.1 | 4 | Nearest neighbour match | 16.2 | NA | NA | NA | NA | NA | NA | NA | NA |
| 1000 | 0.1 | 5 | Nearest neighbour match | 0.1 | NA | NA | NA | NA | NA | NA | NA | NA |
| 100 | 0.1 | 1 | Nearest neighbour match | 67.4 | 0.213431 | 0.001355 | 0.754876 | 0.000825 | 0 | 0 | 66.1 | 0 |
| 100 | 0.1 | 2 | Nearest neighbour match | 29.2 | 0.286568 | 0.002642 | 0.635487 | 0.001787 | 0.100301 | 0.100251 | 28.78636 | 0.100251 |
| 100 | 0.1 | 3 | Nearest neighbour match | 6.2 | NA | NA | NA | NA | NA | NA | NA | NA |
| 100 | 0.1 | 4 | Nearest neighbour match | 0.9 | NA | NA | NA | NA | NA | NA | NA | NA |
| 100 | 0.1 | 5 | Nearest neighbour match | 0 | NA | NA | NA | NA | NA | NA | NA | NA |
| 1.00E+05 | 0.1 | 1 | Caliper match | 100 | 0.002304 | 3.40E-05 | 0.026963 | 2.08E-05 | 100 | 0 | 96 | 0 |
| 1.00E+05 | 0.1 | 2 | Caliper match | 100 | -0.05242 | 5.18E-05 | 0.06083 | 4.16E-05 | 100 | 0 | 87.4 | 0 |
| 1.00E+05 | 0.1 | 3 | Caliper match | 100 | -0.13858 | 0.000101 | 0.145884 | 8.98E-05 | 100 | 0 | 72.7 | 0 |
| 1.00E+05 | 0.1 | 4 | Caliper match | 100 | -0.20208 | 0.000212 | 0.237302 | 0.000171 | 97.3 | 0.512552 | 87.3 | 0.512552 |
| 1.00E+05 | 0.1 | 5 | Caliper match | 42.5 | 0.107365 | 0.001568 | 0.569903 | 0.000848 | 0 | 0 | 42.5 | 0 |
| 10000 | 0.1 | 1 | Caliper match | 100 | -0.00212 | 0.0001 | 0.079867 | 6.06E-05 | 100 | 0 | 97.7 | 0 |
| 10000 | 0.1 | 2 | Caliper match | 100 | -0.05135 | 0.000165 | 0.138374 | 0.000103 | 100 | 0 | 96.5 | 0 |
| 10000 | 0.1 | 3 | Caliper match | 100 | -0.15288 | 0.000322 | 0.278344 | 0.000223 | 77.6 | 1.318423 | 95.3 | 1.318423 |
| 10000 | 0.1 | 4 | Caliper match | 96.2 | -0.22753 | 0.000716 | 0.572583 | 0.000462 | 0.7 | 0.263647 | 95.7 | 0.263647 |
| 10000 | 0.1 | 5 | Caliper match | 0.9 | NA | NA | NA | NA | NA | NA | NA | NA |

**Supp table 2b: Simulation results for exposure probability of 0.1 (continued from 2a, continues in 2c)**

| n | Exposure probability | PS overlap scenario | Method | Convergence (%) | Bias | SE bias | Absolute error | SE absolute error | Power (%) | SE Power | Coverage (%) | SE coverage |
|---|---|---|---|---|---|---|---|---|---|---|---|---|
| 1000 | 0.1 | 1 | Caliper match | 100 | -0.00785 | 0.000358 | 0.281603 | 0.00022 | 70.1 | 1.447753 | 96.9 | 1.447753 |
| 1000 | 0.1 | 2 | Caliper match | 99.7 | -0.10773 | 0.000586 | 0.458602 | 0.000378 | 5.8 | 0.739162 | 98.5 | 0.739162 |
| 1000 | 0.1 | 3 | Caliper match | 72.2 | 0.08761 | 0.001073 | 0.640345 | 0.000615 | 0 | 0 | 71.9 | 0 |
| 1000 | 0.1 | 4 | Caliper match | 13.4 | NA | NA | NA | NA | NA | NA | NA | NA |
| 1000 | 0.1 | 5 | Caliper match | 0 | NA | NA | NA | NA | NA | NA | NA | NA |
| 100 | 0.1 | 1 | Caliper match | 59.8 | 0.211592 | 0.001479 | 0.752363 | 0.000853 | 0 | 0 | 59.7 | 0 |
| 100 | 0.1 | 2 | Caliper match | 23.7 | NA | NA | NA | NA | NA | NA | NA | NA |
| 100 | 0.1 | 3 | Caliper match | 3.7 | NA | NA | NA | NA | NA | NA | NA | NA |
| 100 | 0.1 | 4 | Caliper match | 0.5 | NA | NA | NA | NA | NA | NA | NA | NA |
| 100 | 0.1 | 5 | Caliper match | 0 | NA | NA | NA | NA | NA | NA | NA | NA |
| 1.00E+05 | 0.1 | 1 | IPTW | 100 | -0.02878 | 2.95E-05 | 0.033615 | 2.39E-05 | 100 | 0 | 87.8 | 0 |
| 1.00E+05 | 0.1 | 2 | IPTW | 100 | 0.012763 | 6.81E-05 | 0.05423 | 4.31E-05 | 100 | 0 | 93.7 | 0 |
| 1.00E+05 | 0.1 | 3 | IPTW | 100 | 0.076698 | 0.000244 | 0.205244 | 0.000152 | 96.4 | 0.589101 | 85.2 | 0.589101 |
| 1.00E+05 | 0.1 | 4 | IPTW | 100 | 0.437139 | 0.000734 | 0.698854 | 0.000491 | 15.8 | 1.153412 | 65.3 | 1.153412 |
| 1.00E+05 | 0.1 | 5 | IPTW | 96.4 | 5.247144 | 0.003104 | 5.452763 | 0.002695 | 76.4 | 1.342773 | 11.7 | 1.342773 |
| 10000 | 0.1 | 1 | IPTW | 100 | -0.0302 | 9.44E-05 | 0.079833 | 5.86E-05 | 100 | 0 | 95.8 | 0 |
| 10000 | 0.1 | 2 | IPTW | 100 | 0.026554 | 0.000204 | 0.161412 | 0.000128 | 99.2 | 0.281709 | 93.1 | 0.281709 |
| 10000 | 0.1 | 3 | IPTW | 100 | 0.263692 | 0.000652 | 0.553033 | 0.000434 | 26.8 | 1.400628 | 76.9 | 1.400628 |
| 10000 | 0.1 | 4 | IPTW | 99.8 | 1.242227 | 0.001565 | 1.603137 | 0.00119 | 19.6 | 1.255325 | 54.3 | 1.255325 |
| 10000 | 0.1 | 5 | IPTW | 27.4 | 4.640378 | 0.010549 | 5.066658 | 0.007481 | 21.9 | 1.307819 | 3.5 | 1.307819 |
| 1000 | 0.1 | 1 | IPTW | 100 | 0.00733 | 0.000331 | 0.261345 | 0.000202 | 81.8 | 1.220147 | 95.6 | 1.220147 |
| 1000 | 0.1 | 2 | IPTW | 100 | 0.149046 | 0.00066 | 0.52233 | 0.00043 | 16.8 | 1.182269 | 90.3 | 1.182269 |
| 1000 | 0.1 | 3 | IPTW | 93.7 | 0.92096 | 0.001796 | 1.468067 | 0.001318 | 13 | 1.063485 | 62.2 | 1.063485 |
| 1000 | 0.1 | 4 | IPTW | 50.8 | 1.744638 | 0.004728 | 2.473204 | 0.003229 | 16.5 | 1.173776 | 22.9 | 1.173776 |
| 1000 | 0.1 | 5 | IPTW | 2.5 | NA | NA | NA | NA | NA | NA | NA | NA |

**Supp table 2c: Simulation results for exposure probability of 0.1 (continued from 2b, continues in 2d)**

| n | Exposure probability | PS overlap scenario | Method | Convergence (%) | Bias | SE bias | Absolute error | SE absolute error | Power (%) | SE Power | Coverage (%) | SE coverage |
|---|---|---|---|---|---|---|---|---|---|---|---|---|
| 100 | 0.1 | 1 | IPTW | 89.4 | 0.225603 | 0.001342 | 0.963315 | 0.000838 | 0.7 | 0.263647 | 80.7 | 0.263647 |
| 100 | 0.1 | 2 | IPTW | 57.8 | -0.02142 | 0.002416 | 1.117318 | 0.001448 | 0.902708 | 0.299542 | 50.75226 | 0.299542 |
| 100 | 0.1 | 3 | IPTW | 22.6 | NA | NA | NA | NA | NA | NA | NA | NA |
| 100 | 0.1 | 4 | IPTW | 5.2 | NA | NA | NA | NA | NA | NA | NA | NA |
| 100 | 0.1 | 5 | IPTW | 0 | NA | NA | NA | NA | NA | NA | NA | NA |
| 1.00E+05 | 0.1 | 1 | Logistic regression | 100 | -9.81E-06 | 2.04E-05 | 0.016516 | 1.20E-05 | 100 | 0 | 94.7 | 0 |
| 1.00E+05 | 0.1 | 2 | Logistic regression | 100 | 0.000273 | 2.65E-05 | 0.021145 | 1.59E-05 | 100 | 0 | 94.7 | 0 |
| 1.00E+05 | 0.1 | 3 | Logistic regression | 100 | 0.000797 | 4.25E-05 | 0.033952 | 2.55E-05 | 100 | 0 | 93.8 | 0 |
| 1.00E+05 | 0.1 | 4 | Logistic regression | 100 | 0.004359 | 7.37E-05 | 0.058738 | 4.46E-05 | 100 | 0 | 94.8 | 0 |
| 1.00E+05 | 0.1 | 5 | Logistic regression | 96.4 | 0.042781 | 0.000322 | 0.246071 | 0.000202 | 80.1 | 1.262533 | 92.6 | 1.262533 |
| 10000 | 0.1 | 1 | Logistic regression | 100 | -0.00238 | 5.96E-05 | 0.04782 | 3.55E-05 | 100 | 0 | 96.2 | 0 |
| 10000 | 0.1 | 2 | Logistic regression | 100 | 0.003365 | 8.24E-05 | 0.067052 | 4.79E-05 | 100 | 0 | 96 | 0 |
| 10000 | 0.1 | 3 | Logistic regression | 100 | 0.009326 | 0.000131 | 0.104426 | 7.97E-05 | 100 | 0 | 94.9 | 0 |
| 10000 | 0.1 | 4 | Logistic regression | 99.8 | 0.033202 | 0.000253 | 0.192134 | 0.000168 | 93.9 | 0.756829 | 95.4 | 0.756829 |
| 10000 | 0.1 | 5 | Logistic regression | 27.4 | -0.66785 | 0.000682 | 0.66799 | 0.00068 | 0 | 0 | 23.9 | 0 |
| 1000 | 0.1 | 1 | Logistic regression | 100 | 0.01625 | 0.000205 | 0.161587 | 0.000127 | 99.6 | 0.1996 | 94.8 | 0.1996 |
| 1000 | 0.1 | 2 | Logistic regression | 100 | 0.013902 | 0.000266 | 0.208961 | 0.000166 | 93.9 | 0.756829 | 95.8 | 0.756829 |
| 1000 | 0.1 | 3 | Logistic regression | 93.7 | 0.049017 | 0.000422 | 0.313274 | 0.000262 | 54 | 1.576071 | 90.1 | 1.576071 |
| 1000 | 0.1 | 4 | Logistic regression | 50.8 | -0.33885 | 0.000682 | 0.395634 | 0.00055 | 0.7 | 0.263647 | 46.1 | 0.263647 |
| 1000 | 0.1 | 5 | Logistic regression | 2.7 | NA | NA | NA | NA | NA | NA | NA | NA |
| 100 | 0.1 | 1 | Logistic regression | 89.4 | -0.02653 | 0.000728 | 0.520035 | 0.000439 | 4.4 | 0.648568 | 85 | 0.648568 |
| 100 | 0.1 | 2 | Logistic regression | 57.8 | -0.36494 | 0.001063 | 0.578257 | 0.000725 | 1.2 | 0.344325 | 52.2 | 0.344325 |
| 100 | 0.1 | 3 | Logistic regression | 22.9 | NA | NA | NA | NA | NA | NA | NA | NA |
| 100 | 0.1 | 4 | Logistic regression | 6.2 | NA | NA | NA | NA | NA | NA | NA | NA |
| 100 | 0.1 | 5 | Logistic regression | 0 | NA | NA | NA | NA | NA | NA | NA | NA |

**Supp table 2d: Simulation results for exposure probability of 0.1 (continued from 2c)**

| n | Exposure probability | PS overlap scenario | Method | Convergence (%) | Bias | SE bias | Absolute error | SE absolute error | Power (%) | SE Power | Coverage (%) | SE coverage |
|---|---|---|---|---|---|---|---|---|---|---|---|---|
| 1.00E+05 | 0.5 | 1 | PS covariate | 100 | -0.03559 | 1.71E-05 | 0.035727 | 1.68E-05 | 100 | 0 | 50.8 | 0 |
| 1.00E+05 | 0.5 | 2 | PS covariate | 100 | -0.13116 | 2.53E-05 | 0.131164 | 2.53E-05 | 100 | 0 | 0 | 0 |
| 1.00E+05 | 0.5 | 3 | PS covariate | 100 | -0.1075 | 3.90E-05 | 0.107533 | 3.89E-05 | 100 | 0 | 31 | 0 |
| 1.00E+05 | 0.5 | 4 | PS covariate | 100 | -0.01191 | 6.72E-05 | 0.0544 | 4.12E-05 | 100 | 0 | 98.1 | 0 |
| 1.00E+05 | 0.5 | 5 | PS covariate | 8.8 | NA | NA | NA | NA | NA | NA | NA | NA |
| 10000 | 0.5 | 1 | PS covariate | 100 | -0.03726 | 5.29E-05 | 0.052802 | 3.74E-05 | 100 | 0 | 93.4 | 0 |
| 10000 | 0.5 | 2 | PS covariate | 100 | -0.13196 | 8.04E-05 | 0.135177 | 7.49E-05 | 100 | 0 | 67.8 | 0 |
| 10000 | 0.5 | 3 | PS covariate | 100 | -0.10398 | 0.00013 | 0.136964 | 9.40E-05 | 100 | 0 | 89.6 | 0 |
| 10000 | 0.5 | 4 | PS covariate | 98.7 | -0.00621 | 0.000205 | 0.160355 | 0.000125 | 98.1 | 0.431729 | 97 | 0.431729 |
| 10000 | 0.5 | 5 | PS covariate | 31 | -0.31753 | 0.01876 | 0.992339 | 0.018513 | 0.5 | 0.223047 | 30.4 | 0.223047 |
| 1000 | 0.5 | 1 | PS covariate | 100 | -0.03414 | 0.000166 | 0.135533 | 0.000101 | 100 | 0 | 97.3 | 0 |
| 1000 | 0.5 | 2 | PS covariate | 100 | -0.11593 | 0.00025 | 0.220191 | 0.000165 | 95.2 | 0.675988 | 95.8 | 0.675988 |
| 1000 | 0.5 | 3 | PS covariate | 100 | -0.10236 | 0.000417 | 0.336131 | 0.000267 | 41.8 | 1.559731 | 96.3 | 1.559731 |
| 1000 | 0.5 | 4 | PS covariate | 90.1 | -0.09989 | 0.000945 | 0.634292 | 0.00064 | 0.2 | 0.14128 | 87.9 | 0.14128 |
| 1000 | 0.5 | 5 | PS covariate | 9.3 | NA | NA | NA | NA | NA | NA | NA | NA |
| 100 | 0.5 | 1 | PS covariate | 100 | -0.04522 | 0.000569 | 0.440748 | 0.000362 | 1.2 | 0.344325 | 98.3 | 0.344325 |
| 100 | 0.5 | 2 | PS covariate | 95.8 | -0.24232 | 0.001052 | 0.786767 | 0.000704 | 0 | 0 | 93.9 | 0 |
| 100 | 0.5 | 3 | PS covariate | 62.3 | -1.71912 | 0.013928 | 2.1133 | 0.013787 | 0.1001 | 0.10005 | 61.06106 | 0.10005 |
| 100 | 0.5 | 4 | PS covariate | 20.9 | NA | NA | NA | NA | NA | NA | NA | NA |
| 100 | 0.5 | 5 | PS covariate | 0 | NA | NA | NA | NA | NA | NA | NA | NA |
| 1.00E+05 | 0.5 | 1 | Nearest neighbour match | 100 | -0.01783 | 2.41E-05 | 0.02449 | 1.73E-05 | 100 | 0 | 75.3 | 0 |
| 1.00E+05 | 0.5 | 2 | Nearest neighbour match | 100 | -0.08936 | 3.59E-05 | 0.089558 | 3.54E-05 | 100 | 0 | 18.4 | 0 |
| 1.00E+05 | 0.5 | 3 | Nearest neighbour match | 100 | -0.18261 | 8.42E-05 | 0.184139 | 8.08E-05 | 100 | 0 | 17.8 | 0 |
| 1.00E+05 | 0.5 | 4 | Nearest neighbour match | 100 | -0.29783 | 0.000276 | 0.356013 | 0.000195 | 99.3 | 0.263647 | 25.6 | 0.263647 |
| 1.00E+05 | 0.5 | 5 | Nearest neighbour match | 79.9 | -0.93885 | 0.001231 | 1.173077 | 0.00086 | 4.3 | 0.64149 | 63.1 | 0.64149 |

**Supp table 3a: Simulation results for exposure probability of 0.5 (continues in 3b)**

| n | Exposure probability | PS overlap scenario | Method | Convergence (%) | Bias | SE bias | Absolute error | SE absolute error | Power (%) | SE Power | Coverage (%) | SE coverage |
|---|---|---|---|---|---|---|---|---|---|---|---|---|
| 10000 | 0.5 | 1 | Nearest neighbour match | 100 | -0.01903 | 7.39E-05 | 0.06022 | 4.69E-05 | 100 | 0 | 86.5 | 0 |
| 10000 | 0.5 | 2 | Nearest neighbour match | 100 | -0.09097 | 0.000116 | 0.120106 | 8.58E-05 | 100 | 0 | 77.1 | 0 |
| 10000 | 0.5 | 3 | Nearest neighbour match | 100 | -0.20595 | 0.000263 | 0.277617 | 0.000186 | 97.7 | 0.474036 | 66.1 | 0.474036 |
| 10000 | 0.5 | 4 | Nearest neighbour match | 100 | -0.47674 | 0.000636 | 0.634051 | 0.000479 | 38.4 | 1.537999 | 70.2 | 1.537999 |
| 10000 | 0.5 | 5 | Nearest neighbour match | 11 | NA | NA | NA | NA | NA | NA | NA | NA |
| 1000 | 0.5 | 1 | Nearest neighbour match | 100 | -0.02525 | 0.00024 | 0.186378 | 0.000153 | 99.2 | 0.281709 | 86.8 | 0.281709 |
| 1000 | 0.5 | 2 | Nearest neighbour match | 100 | -0.13781 | 0.000376 | 0.315417 | 0.000247 | 79 | 1.288022 | 89.1 | 1.288022 |
| 1000 | 0.5 | 3 | Nearest neighbour match | 98 | -0.3375 | 0.000781 | 0.651652 | 0.000535 | 9.5 | 0.927227 | 89.8 | 0.927227 |
| 1000 | 0.5 | 4 | Nearest neighbour match | 45.6 | 0.151331 | 0.002158 | 0.729368 | 0.001485 | 0.6 | 0.244213 | 43.7 | 0.244213 |
| 1000 | 0.5 | 5 | Nearest neighbour match | 0.1 | NA | NA | NA | NA | NA | NA | NA | NA |
| 100 | 0.5 | 1 | Nearest neighbour match | 97.4 | -0.10068 | 0.000821 | 0.637978 | 0.000505 | 2.2 | 0.463853 | 89.3 | 0.463853 |
| 100 | 0.5 | 2 | Nearest neighbour match | 72.5 | 0.076048 | 0.00123 | 0.720246 | 0.000732 | 0.4 | 0.1996 | 70 | 0.1996 |
| 100 | 0.5 | 3 | Nearest neighbour match | 22.2 | NA | NA | NA | NA | NA | NA | NA | NA |
| 100 | 0.5 | 4 | Nearest neighbour match | 2 | NA | NA | NA | NA | NA | NA | NA | NA |
| 100 | 0.5 | 5 | Nearest neighbour match | 0 | NA | NA | NA | NA | NA | NA | NA | NA |
| 1.00E+05 | 0.5 | 1 | Caliper match | 100 | -0.00953 | 2.07E-05 | 0.01858 | 1.32E-05 | 100 | 0 | 95.5 | 0 |
| 1.00E+05 | 0.5 | 2 | Caliper match | 100 | -0.0929 | 3.10E-05 | 0.092941 | 3.09E-05 | 100 | 0 | 19.7 | 0 |
| 1.00E+05 | 0.5 | 3 | Caliper match | 100 | -0.16837 | 6.14E-05 | 0.168492 | 6.10E-05 | 100 | 0 | 26.2 | 0 |
| 1.00E+05 | 0.5 | 4 | Caliper match | 100 | -0.22996 | 0.000142 | 0.235591 | 0.000132 | 100 | 0 | 64.1 | 0 |
| 1.00E+05 | 0.5 | 5 | Caliper match | 74.4 | -0.07988 | 0.000986 | 0.60845 | 0.00056 | 0 | 0 | 74.2 | 0 |
| 10000 | 0.5 | 1 | Caliper match | 100 | -0.01288 | 6.62E-05 | 0.054229 | 4.01E-05 | 100 | 0 | 96.9 | 0 |
| 10000 | 0.5 | 2 | Caliper match | 100 | -0.09723 | 0.0001 | 0.113426 | 8.13E-05 | 100 | 0 | 87.1 | 0 |
| 10000 | 0.5 | 3 | Caliper match | 100 | -0.16976 | 0.000205 | 0.218036 | 0.000153 | 98.9 | 0.329833 | 88.5 | 0.329833 |
| 10000 | 0.5 | 4 | Caliper match | 99.9 | -0.27877 | 0.000488 | 0.436021 | 0.000354 | 29.1 | 1.436381 | 94.9 | 1.436381 |
| 10000 | 0.5 | 5 | Caliper match | 2.9 | NA | NA | NA | NA | NA | NA | NA | NA |

**Supp table 3b: Simulation results for exposure probability of 0.5 (continued from 3a, continues in 3c)**

| n | Exposure probability | PS overlap scenario | Method | Convergence (%) | Bias | SE bias | Absolute error | SE absolute error | Power (%) | SE Power | Coverage (%) | SE coverage |
|---|---|---|---|---|---|---|---|---|---|---|---|---|
| 1000 | 0.5 | 1 | Caliper match | 100 | -0.01122 | 0.000229 | 0.179753 | 0.000142 | 99 | 0.314643 | 96.6 | 0.314643 |
| 1000 | 0.5 | 2 | Caliper match | 100 | -0.09563 | 0.000347 | 0.279412 | 0.000227 | 71.5 | 1.427498 | 96.5 | 1.427498 |
| 1000 | 0.5 | 3 | Caliper match | 97.8 | -0.18793 | 0.000751 | 0.591511 | 0.000484 | 0.7 | 0.263647 | 96.5 | 0.263647 |
| 1000 | 0.5 | 4 | Caliper match | 30.8 | 0.151155 | 0.002122 | 0.571359 | 0.001137 | 0 | 0 | 30.8 | 0 |
| 1000 | 0.5 | 5 | Caliper match | 0 | NA | NA | NA | NA | NA | NA | NA | NA |
| 100 | 0.5 | 1 | Caliper match | 81.6 | 0.101127 | 0.001072 | 0.707704 | 0.000642 | 0 | 0 | 80.9 | 0 |
| 100 | 0.5 | 2 | Caliper match | 48.4 | 0.173271 | 0.001677 | 0.691049 | 0.000947 | 0 | 0 | 48.4 | 0 |
| 100 | 0.5 | 3 | Caliper match | 7.7 | NA | NA | NA | NA | NA | NA | NA | NA |
| 100 | 0.5 | 4 | Caliper match | 0.5 | NA | NA | NA | NA | NA | NA | NA | NA |
| 100 | 0.5 | 5 | Caliper match | 0 | NA | NA | NA | NA | NA | NA | NA | NA |
| 1.00E+05 | 0.5 | 1 | IPTW | 100 | 6.32E-05 | 1.59E-05 | 0.012734 | 9.50E-06 | 100 | 0 | 97.7 | 0 |
| 1.00E+05 | 0.5 | 2 | IPTW | 100 | -0.00164 | 2.94E-05 | 0.023238 | 1.81E-05 | 100 | 0 | 96.7 | 0 |
| 1.00E+05 | 0.5 | 3 | IPTW | 100 | -0.00382 | 0.000143 | 0.112017 | 8.92E-05 | 99.6 | 0.1996 | 92.4 | 0.1996 |
| 1.00E+05 | 0.5 | 4 | IPTW | 100 | 0.19742 | 0.000576 | 0.482599 | 0.00037 | 46.1 | 1.576322 | 72.8 | 1.576322 |
| 1.00E+05 | 0.5 | 5 | IPTW | 100 | 4.582742 | 0.002293 | 4.727789 | 0.001977 | 79.6 | 1.2743 | 10.7 | 1.2743 |
| 10000 | 0.5 | 1 | IPTW | 100 | -0.00156 | 4.92E-05 | 0.039781 | 2.89E-05 | 100 | 0 | 98.5 | 0 |
| 10000 | 0.5 | 2 | IPTW | 100 | 0.004651 | 9.24E-05 | 0.072758 | 5.71E-05 | 100 | 0 | 96.8 | 0 |
| 10000 | 0.5 | 3 | IPTW | 100 | 0.07874 | 0.000388 | 0.307416 | 0.000249 | 82 | 1.214907 | 84.8 | 1.214907 |
| 10000 | 0.5 | 4 | IPTW | 100 | 0.801685 | 0.001135 | 1.154034 | 0.000773 | 13.2 | 1.070402 | 53.4 | 1.070402 |
| 10000 | 0.5 | 5 | IPTW | 64.8 | 5.905272 | 0.003264 | 5.986807 | 0.002889 | 58.7 | 1.55702 | 4 | 1.55702 |
| 1000 | 0.5 | 1 | IPTW | 100 | 0.00359 | 0.000153 | 0.123231 | 9.06E-05 | 100 | 0 | 98.4 | 0 |
| 1000 | 0.5 | 2 | IPTW | 100 | 0.04 | 0.000289 | 0.225889 | 0.000184 | 89.6 | 0.965319 | 96.1 | 0.965319 |
| 1000 | 0.5 | 3 | IPTW | 100 | 0.429436 | 0.000914 | 0.784555 | 0.000636 | 7.3 | 0.822624 | 74.6 | 0.822624 |
| 1000 | 0.5 | 4 | IPTW | 90.2 | 2.281712 | 0.002063 | 2.544733 | 0.001641 | 40.2 | 1.550471 | 33.3 | 1.550471 |
| 1000 | 0.5 | 5 | IPTW | 9.5 | NA | NA | NA | NA | NA | NA | NA | NA |

**Supp table 3c: Simulation results for exposure probability of 0.5 (continued from 3c, continues in 3d)**

| n | Exposure probability | PS overlap scenario | Method | Convergence (%) | Bias | SE bias | Absolute error | SE absolute error | Power (%) | SE Power | Coverage (%) | SE coverage |
|---|---|---|---|---|---|---|---|---|---|---|---|---|
| 100 | 0.5 | 1 | IPTW | 100 | 0.014764 | 0.00055 | 0.419714 | 0.000356 | 3.9 | 0.612201 | 97.4 | 0.612201 |
| 100 | 0.5 | 2 | IPTW | 95.8 | 0.240202 | 0.001013 | 0.783595 | 0.000647 | 0 | 0 | 89 | 0 |
| 100 | 0.5 | 3 | IPTW | 62.6 | 0.774853 | 0.002299 | 1.344331 | 0.001484 | 1.101101 | 0.330162 | 48.84885 | 0.330162 |
| 100 | 0.5 | 4 | IPTW | 21.4 | NA | NA | NA | NA | NA | NA | NA | NA |
| 100 | 0.5 | 5 | IPTW | 0 | NA | NA | NA | NA | NA | NA | NA | NA |
| 1.00E+05 | 0.5 | 1 | Logistic regression | 100 | 0.000654 | 1.44E-05 | 0.011603 | 8.53E-06 | 100 | 0 | 94.5 | 0 |
| 1.00E+05 | 0.5 | 2 | Logistic regression | 100 | 6.37E-05 | 1.72E-05 | 0.013656 | 1.05E-05 | 100 | 0 | 94.1 | 0 |
| 1.00E+05 | 0.5 | 3 | Logistic regression | 100 | -0.00012 | 2.50E-05 | 0.019946 | 1.51E-05 | 100 | 0 | 94.6 | 0 |
| 1.00E+05 | 0.5 | 4 | Logistic regression | 100 | 0.002124 | 4.28E-05 | 0.034085 | 2.60E-05 | 100 | 0 | 94.9 | 0 |
| 1.00E+05 | 0.5 | 5 | Logistic regression | 100 | 0.026317 | 0.000181 | 0.140248 | 0.000118 | 99.1 | 0.298647 | 94.7 | 0.298647 |
| 10000 | 0.5 | 1 | Logistic regression | 100 | 0.000144 | 4.44E-05 | 0.035327 | 2.68E-05 | 100 | 0 | 94 | 0 |
| 10000 | 0.5 | 2 | Logistic regression | 100 | 0.000263 | 5.33E-05 | 0.042899 | 3.17E-05 | 100 | 0 | 95.1 | 0 |
| 10000 | 0.5 | 3 | Logistic regression | 100 | 0.003195 | 7.94E-05 | 0.064151 | 4.68E-05 | 100 | 0 | 95.6 | 0 |
| 10000 | 0.5 | 4 | Logistic regression | 100 | 0.012328 | 0.000134 | 0.104885 | 8.36E-05 | 99.9 | 0.09995 | 95.9 | 0.09995 |
| 10000 | 0.5 | 5 | Logistic regression | 64.8 | -0.16349 | 0.000365 | 0.220438 | 0.000285 | 41.2 | 1.556458 | 60.1 | 1.556458 |
| 1000 | 0.5 | 1 | Logistic regression | 100 | 0.001655 | 0.000138 | 0.109866 | 8.37E-05 | 100 | 0 | 95.4 | 0 |
| 1000 | 0.5 | 2 | Logistic regression | 100 | 0.012929 | 0.000171 | 0.137438 | 0.000102 | 100 | 0 | 95.7 | 0 |
| 1000 | 0.5 | 3 | Logistic regression | 100 | 0.024236 | 0.000258 | 0.203358 | 0.000161 | 94.2 | 0.739162 | 96.5 | 0.739162 |
| 1000 | 0.5 | 4 | Logistic regression | 90.2 | 0.045674 | 0.000429 | 0.31047 | 0.00026 | 51.9 | 1.579997 | 86.4 | 1.579997 |
| 1000 | 0.5 | 5 | Logistic regression | 9.9 | NA | NA | NA | NA | NA | NA | NA | NA |
| 100 | 0.5 | 1 | Logistic regression | 100 | 0.001625 | 0.000474 | 0.371008 | 0.000296 | 42.6 | 1.563726 | 93.7 | 1.563726 |
| 100 | 0.5 | 2 | Logistic regression | 94.6 | -0.00433 | 0.000644 | 0.485514 | 0.000389 | 14.4 | 1.110243 | 87.2 | 1.110243 |
| 100 | 0.5 | 3 | Logistic regression | 59.7 | -0.21994 | 0.001148 | 0.579225 | 0.000715 | 2.5 | 0.49371 | 52.9 | 0.49371 |
| 100 | 0.5 | 4 | Logistic regression | 22.1 | NA | NA | NA | NA | NA | NA | NA | NA |
| 100 | 0.5 | 5 | Logistic regression | 1.8 | NA | NA | NA | NA | NA | NA | NA | NA |

**Supp table 3d: Simulation results for exposure probability of 0.5 (continued from 3c)**